\PassOptionsToPackage{hyphens}{url}
\documentclass[sigconf, screen, nonacm]{acmart}

\usepackage{amsmath}
\usepackage{algorithmic}
\usepackage{graphicx}
\usepackage{textcomp}
\usepackage{xcolor}
\usepackage{enumitem}
\usepackage[normalem]{ulem}

\newcommand{\system}{Ambar}

\begin{document}

\title{\system{}: Memory Management for Agentic AI Sandboxes at Scale}

\makeatletter
\newcommand{\paperNames}{}
\newcommand{\paperAuthor}[3][]{%
  \author{#2}%
  \affiliation{\institution{#3}}%
  \g@addto@macro\paperNames{#2#1\enspace}%
}
\newcommand{\utmark}{\textsuperscript{1}}
\newcommand{\metamark}{\textsuperscript{2}}

\newcommand{\utandmetamark}{\textsuperscript{1,2}}

\paperAuthor[\utmark]{Yigit Kemal Can}{The University of Texas at Austin}
\paperAuthor[\metamark]{Pratyush Patel}{Meta}
\paperAuthor[\metamark]{Neha Gholkar}{Meta}
\paperAuthor[\metamark]{Hasan Maruf}{Meta}
\paperAuthor[\metamark]{Apostolos Kokolis}{Meta}
\paperAuthor[\utandmetamark]{Jovan Stojkovic}{The University of Texas at Austin and Meta}

\if@ACM@anonymous\else
  \renewcommand{\@mkauthors@iii}{%
    \global\setbox\mktitle@bx=\vbox{%
      \unvbox\mktitle@bx
      \vskip 13.5pt
      \hbox to \textwidth{\hfil
        {\normalfont\fontsize{10}{12}\selectfont\paperNames\unskip}%
        \hfil}%
      \vskip 3pt
      \hbox to \textwidth{\hfil
        {\@affiliationfont The University of Texas at Austin\utmark, Meta\metamark}%
        \hfil}%
      \bigskip}}
\fi
\makeatother

\renewcommand{\shortauthors}{Y. Can, P. Patel, N. Gholkar, H. Maruf, A. Kokolis, J. Stojkovic}

\begin{abstract}
Agentic AI creates growing demand for datacenter CPUs to
execute tools inside sandboxes, e.g., VMs or containers.
Our characterization of open-source agentic AI benchmarks and
our sandbox workloads 
shows that
\emph{memory capacity} prevents servers from fully utilizing
these CPUs. The cause is a mismatch between CPU and memory
demand: sandboxes execute tools in short bursts, then wait
for LLM responses or user input while their memory remains
stranded.
Servers can use these idle periods to colocate
more agents and improve datacenter efficiency, but server memory capacity limits the number
of agents each can support. To increase effective memory
capacity, idle sandboxes can be compressed or moved to slower memory tiers,
such as CXL memory or SSDs.
However, these
operations and subsequent memory restoration
consume CPU cycles and delay tool execution. Hence,
memory management must increase effective memory capacity while preserving CPU cycles for new
agents and avoiding delays when idle agents resume
execution.

Guided by our insights, we propose \emph{\system{}},
an agent-aware memory manager that jointly manages memory
tiering and compression to support more concurrent agents
per server while limiting execution overhead.
Under memory pressure, \system{} decides which sandboxes
to move to a slower memory tier, compress, or both.
To do that, it predicts each sandbox's idle time and
weighs the local memory it frees against
the time needed to reclaim and restore its memory before tool
execution resumes.
To make reclamation practical even during
short idle periods, \system{} introduces three optimizations.
It combines (i) efficient snapshotting with (ii) hot-page tracking
to reduce state-preservation work and eagerly restore
only the memory that the next tool needs.
To reduce the CPU overhead of these operations, (iii) \system{}
offloads data movement and compression to hardware accelerators.
Compared to conventional agentic sandbox frameworks,
\system{} increases the number of agents per server by
5.9$\times$ and CPU utilization by 5.0$\times$, while limiting
agent P50 and P99 latency increases by up to 5.6\% and
6.3\%, respectively.
\end{abstract}

\maketitle 

\section{Introduction}
\label{sec:introduction}

Agentic AI is becoming an important workload in datacenters~\cite{yang2026architectural,chang2026agentsysbench,liu2026agenticcoding}.
It automates multi-step tasks such as software development~\cite{yang2024sweagent,jimenez2024swebench}, data analytics~\cite{egg2025dabstep,guo2024dsagent}, and scientific research~\cite{lu2024aiscientist}. Unlike standalone LLM inference, agentic AI combines model reasoning with software tools that execute commands and process data~\cite{yao2023react,wu2024autogen,wang2024codeact}.
For example, a coding agent can modify a program, run tests,
and use the results to determine its next action.
Platforms typically execute these tools inside sandboxes,
such as VMs or containers, to isolate each agent from
the host and other agents~\cite{openaisandboxes,wang2025openhands}.
While standalone LLM inference
has focused attention on accelerator performance,
agentic AI brings CPUs back to the forefront of AI system design.
CPUs execute the tools that turn model decisions into
actions, making CPU server efficiency essential
to the overall performance of agentic AI services~\cite{raj2025cpucentric,yang2026architectural}.

However, due to their fragmented execution, agentic workloads do not use CPUs efficiently~\cite{yang2026architectural,zheng2026agentcgroup}. A sandbox
executes a tool invoked by the LLM, returns the result, and waits
for the next LLM response. After the request completes,
it remains idle until the next user input.
We observe these long pauses between tool invocations
in both open-source frameworks and 
our eight additional sandbox workloads. Across workloads, tools execute for fractions
of a second,
LLM calls take several
seconds, and user idle times are minutes to hours.
Hence, a typical sandbox has very low
CPU utilization.

To improve CPU utilization, a server can colocate more agents. While an agent waits, another executes
tools on the available cores. This allows the datacenter
to support more agents with fewer servers.
However, idle sandboxes retain their memory as next tool calls depend on the state from
earlier steps~\cite{chang2026agentsysbench,huang2026dsec}. Hence, \emph{memory capacity} limits the
number of agents that a server can support~\cite{zheng2026agentcgroup}.
We observe
that servers exhaust memory capacity before they fully
utilize their CPUs. On the most densely packed 
hosts that we measure, memory reservations approach full capacity while
more than 90\% of CPU cores remain idle. To use these
available cores for additional agents, we must reclaim
local memory from idle sandboxes.

A simple solution would follow the approach of traditional
datacenter workloads, such as stateless serverless functions~\cite{shahrad2020serverlesswild,oakes2018sock}.
As function invocations are independent, platforms can
remove idle sandboxes and recreate them on
demand. This approach creates two issues
for agentic workloads. First, sandbox initialization is
on the tool's critical path~\cite{du2020catalyzer,ustiugov2021reap,brooker2023containerloading,holmes2026spice}. A request can have
tens of tool--LLM transitions, so repeated initialization
adds substantial overhead to short tool executions. Second, agentic sandboxes
preserve state across steps. Tools may depend
on files, loaded datasets, or live processes from earlier
invocations~\cite{wang2025openhands,chang2026agentsysbench}. Recreating a clean sandbox loses this state.
Hence, we need memory reclamation that preserves sandbox
state and lets tools resume without re-initialization.

A promising approach is to increase server's effective
memory capacity via memory tiering~\cite{weiner2022tmo,maruf2023tpp,lagarcavilla2019far} and compression~\cite{jeong2023compression,karandikar2023cdpu}.
During idle periods, a memory manager can move sandbox
memory to slower tiers, such as CXL-attached memory or SSDs,
or compress it to free local capacity. Then, it must
restore the necessary state before tool execution resumes.
To understand the opportunities and challenges of this
approach, we characterize open-source coding and
data-analytics agents and 
our eight additional sandbox workloads, which cover interactive user sessions, reinforcement learning, and automated evaluation.

Our characterization provides three insights.
First, the duration of an idle period determines which
memory-management operations can finish before tool
execution resumes. LLM inference often provides enough
time to move sandbox memory to CXL and bring it back,
or to compress and decompress it. On the other hand,
agents can wait minutes to hours for user input.
These longer pauses make even slower operations practical,
e.g., offload to SSDs and subsequent restoration. 
Second, the system has an opportunity to 
reduce these overheads.
The median tool invocation accesses only 4\%--6\% of the
sandbox footprint, so the system only needs to bring
those pages back before the tool starts.
This hot working set is predictable, as
tools reuse 90\% of their accessed pages from the latest invocations of the same tool. 
Additionally, hardware accelerators can reduce compression
overhead. For instance, Intel's IAA~\cite{inteliaa} speeds
up compression by 2.5$\times$, on average, over software and reduces
the CPU work required.
Finally, memory-management operations
can severely slow tools if they occur on the critical path.
For example, an 800\,ms CXL round trip would add overhead
equal to 8$\times$ the execution time of a typical 100\,ms tool.
Hence, the system must match reclamation to the available
idle time and prepare the next tool's working set before
execution resumes.

We use these insights to design \system{}, an agent-aware
memory manager that coordinates memory tiering and
compression to support more concurrent agents per server
while limiting execution overhead. Under memory pressure,
\system{} decides which sandboxes to move to a slower
memory tier, compress, or both. For each sandbox, it
predicts the time until the next tool invocation.
It estimates how much local memory migration or compression
would free and the time needed to reclaim and restore
that memory. \system{} compares these estimates to select
operations that fit within the predicted idle period.
To
correct its predictions, \system{} uses LLM hints about upcoming tool calls. If the LLM signals that a tool
invocation is imminent, \system{} 
starts restoration sooner.

To make reclamation practical even during short idle
periods, \system{} introduces three optimizations.
First, its efficient snapshot mechanism tracks state
changes across tool invocations and updates only modified
memory, avoiding repeated snapshots of the full sandbox.
Second, \system{} tracks page reuse to predict which
pages the next tool will access. It prefetches those pages
before the tool starts and fetches other pages only
if the tool accesses them. Third, \system{} coordinates hardware accelerator use
across sandboxes to speed up memory operations and
reduce CPU overhead.

We evaluate \system{} with widely used coding~\cite{yang2024sweagent,jimenez2024swebench} and data-analytics~\cite{egg2025dabstep} agentic AI workloads.
Compared to conventional sandbox management,
\system{} increases the number of agents per server by
5.9$\times$ and CPU utilization by 5.0$\times$, while limiting
agent P50 and P99 latency increases by up to 5.6\% and
6.3\%, respectively. 

This paper makes the following contributions:
\begin{itemize}[leftmargin=*,topsep=2pt,itemsep=1pt]
    \item We characterize 
    open-source agentic workloads and our sandbox workloads
    and show that memory capacity is the primary constraint on the number of concurrent agents per server.

    \item We analyze the implications
    of increasing effective server memory capacity
    through sandbox memory tiering and compression.

    \item We design and evaluate \system{}, an agent-aware
    memory manager that jointly and efficiently manages
    memory tiering and compression to support more
    concurrent agents per server.
\end{itemize}

\section{Background}
\label{sec:background}

\noindent \textbf{Agentic AI workloads.} 
Agentic AI applications perform multi-step tasks by interleaving LLM inference with external tool execution~\cite{yao2023react,wu2024autogen,google2026agentic,microsoftagentic}.   
Tools allow agents to interact with their execution environment by reading files, modifying code, and executing tests~\cite{yang2024sweagent,wang2025openhands,wang2024codeact}.  
LLM inference typically runs on remote GPU servers accessed through API calls, whereas tool execution primarily runs on CPUs.  
In these workloads, a \emph{step} is an LLM invocation and any associated tool execution, while a \emph{turn} is a task initiated by a user or an automated prompt and may involve multiple steps. A \emph{session} consists of multiple turns that share the same execution environment~\cite{openaiagentsessions,zhu2026tracelab,liu2026agenticcoding}. 
Since each step depends on the results of preceding steps, execution alternates between LLM inference and tools, resulting in fragmented CPU activity~\cite{yang2026architectural,zhu2026tracelab,liu2026agenticcoding,raj2025cpucentric,yuan2026workload,kim2026dynamicreasoning}.

\vspace{2pt}
\noindent \textbf{Sandboxes for agents.} 
Agent tools run in isolated sandboxes. This provides a controlled environment with  tools, libraries, and files needed for task execution while isolating agents 
from one another~\cite{wang2025openhands,raj2025cpucentric,zheng2026agentcgroup,openaisandboxes,oakes2018sock,wang2025openhands}. 
They are stateful and persist across  steps and turns, preserving running 
processes, installed dependencies, and  workspace  changes~\cite{chang2026agentsysbench,liu2026agenticcoding,wang2025openhands}. 
Different sandbox mechanisms: containers~\cite{docker,podman}, virtual machines (VMs)~\cite{kivity2007kvm,bellard2005qemu}, or lightweight microVMs~\cite{cloudhypervisor,agache2020firecracker}, 
have trade-offs in performance and isolation.

\vspace{2pt}
\noindent \textbf{Memory hierarchy in datacenters.} 
An approach to address memory capacity constraints is to increase effective capacity through memory tiering. 
Datacenter memory hierarchies combine local DRAM with higher-capacity, lower-performance memory and storage tiers that offer different  characteristics~\cite{dulloor2016datatiering,weiner2022tmo,lee2023memtis,duraisamy2023tiering,xiang2024nomad,zhong2024cxltiering}. 
Solid-state drives (SSDs) provide a 
high-capacity storage tier.
However, accessing data from SSDs requires explicit I/O operations, causing higher access latency~\cite{chen2009ssd,dirik2009ssd,weiner2022tmo,haas2023nvme,ibmssd,pavlovic2024ssd}.  
Between the local DRAM and SSDs, Compute Express Link (CXL) protocol can be used to attach additional DRAM DIMMs. 
Such additional memory 
preserves a byte-addressable load/store interface, offers lower latency than SSDs, but still has higher latency and lower bandwidth than local DRAM~\cite{alverti2025cxlfork,cho2024coaxial,kim2024damon,kim2023smt,li2023pond,liu2024nearcxl,zeng2025samsungcxl}. 

\vspace{2pt}
\noindent \textbf{Memory compression in datacenters.} 
Another approach to increase effective memory capacity is memory compression, which reduces the memory footprint by storing data in a compressed representation~\cite{jeong2023compression,lagarcavilla2019far,choukse2018compresso,weiner2022tmo}. 
Software compression algorithms, such as LZ4~\cite{lz4software} and Zstandard (zstd)~\cite{metazstandard}, offer trade-offs between compression ratio and computational overhead. 
These algorithms consume CPU cycles that could otherwise be used for application execution. 
Dedicated hardware accelerators can reduce this overhead by offloading (de)compression operations from CPU cores. Examples include Intel's In-Memory Analytics Accelerator (IAA)~\cite{inteliaa,kang2025iaa}, Intel QuickAssist Technology (QAT)~\cite{intelqat,xia2025qat}, and IBM's Nest Accelerator (NX)~\cite{abali2020compression}. 
Alternatively, inline hardware compression engines can be integrated into memory subsystems, including CXL memory expanders, to (de)compress directly in the memory subsystem~\cite{chauhan2023cxlexpander}.
\section{Opportunities and Challenges of Memory Management for Agentic Sandboxes}
\label{sec:characterization}

To understand the opportunities and challenges of managing
memory of agentic sandboxes, we characterize open-source
agentic benchmarks on our servers and our eight additional sandbox workloads. We first analyze how sandboxes
execute and use resources (\S~\ref{sec:char-workload}).
Then, we quantify the overhead of memory tiering
(\S~\ref{sec:char-tiering}) and memory compression
(\S~\ref{sec:char-compression}). Here, we describe our
setup and present our main observations.

\vspace{2pt}
\noindent \textbf{Open-source workloads.}
We run two widely used agentic workloads. For coding, we run
mini-SWE-agent~\cite{yang2024sweagent} on 100 tasks from
SWE-bench Verified~\cite{jimenez2024swebench}. In each task,
the agent resolves a GitHub issue by editing a Python
repository and running its tests. For data analytics, we run
the official DABStep agent~\cite{egg2025dabstep} on 100 tasks.
In each task, the agent writes and executes Python code to
answer a question about a set of payment datasets. Both agents
use Qwen3-235B-A22B-Instruct~\cite{qwen3}, which we
serve with vLLM~\cite{vllm} on 8 NVIDIA A100 GPUs.
Each task runs in its own sandbox: either a Docker
container~\cite{docker} or a Cloud Hypervisor
microVM~\cite{cloudhypervisor}.
We run the sandboxes on a server with a 96-core AMD Turin processor, 128\,GB of
local DRAM, and 128\,GB CXL-attached memory.
For compression experiments, we also use an Intel Xeon Gold 6550C server with
on-chip DSA and IAA accelerators.
We record the duration of every LLM call and tool call, and
sample the CPU utilization and memory usage of each sandbox at
sub-second granularity.
As our setup has no human users, we take the idle time distributions between user turns from two public
production traces: GitHub Copilot traces from
Azure~\cite{liu2026agenticcoding} and FreeInference~\cite{nixon2026requests}.

\vspace{2pt}
\noindent \textbf{Our sandbox workloads.}
We also analyze our eight additional sandbox workloads, which we call A--H. Four workloads are interactive, i.e., a human drives the work: 
VMs that execute code (A), VMs that run a
code interpreter (B), development containers for coding
agents (C), and containers for data-analysis agents (D). The
other four run reinforcement learning (RL) and evaluation jobs:
VMs shared by many RL and evaluation tenants (E), VMs that each
execute a single piece of code (F), VMs for browsing agents (G),
and containers for RL and evaluation (H). Hence, the workloads
cover both VMs and containers, and both human-driven and
machine-driven agents. For each workload, we collect one day of
measurements: per-sandbox CPU and memory usage, sandbox
lifetimes, and logs of tool and LLM calls. We report CPU and
memory usage relative to the sandbox's allocation, i.e., the
resources that the scheduler reserves for the sandbox on its
host.

\subsection{Workload Properties}
\label{sec:char-workload}

\begin{figure}[t]
    \vspace{0mm}
    \centering
    \includegraphics[width=\linewidth]{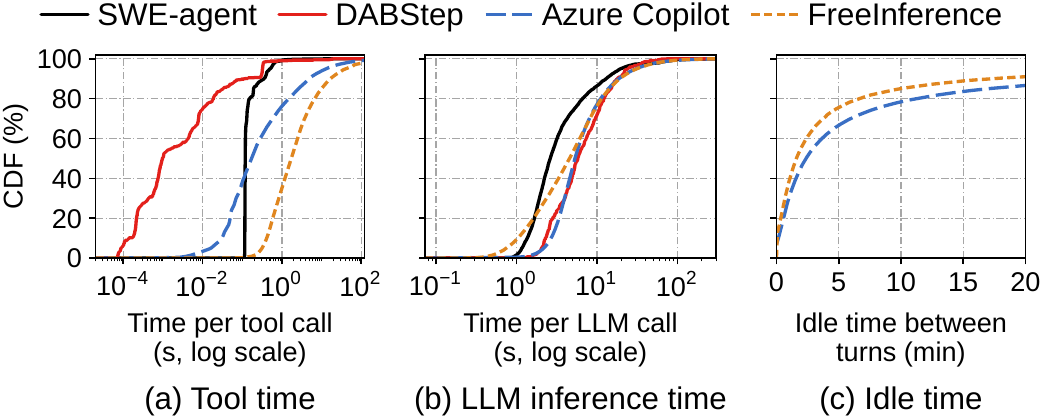}
    \vspace{-5mm}
    \caption{Tool time, LLM inference time, and user idle time distributions across open-source workloads and public traces.}
    \label{fig:time-dist}
    \vspace{-4mm}
\end{figure}

\vspace{2pt}
\noindent \textbf{Tool, LLM, and idle times.}
Figure~\ref{fig:time-dist} shows the distributions of tool
time, LLM inference time, and idle time between user turns in
the open-source workloads and public traces.
Tool calls are short: the median tool
call takes 119\,ms in SWE-agent and only 1.0\,ms in DABStep, and
99\% of the tool calls in both agents finish within 1\,s.
LLM calls are an order of magnitude longer: the median LLM call
takes 2.6\,s in SWE-agent and 5.8\,s in DABStep, and the
P90 call takes 12.9\,s and 17.7\,s, respectively.
Since a task has many steps (32 and 5 for median SWE-agent and DABStep tasks, respectively), the sandbox waits for the LLM many times within each task.
After the task completes, the sandbox remains idle until the next user turn, which takes even longer: the median idle time is 2.4\,min in the Azure Copilot traces and 1.6\,min in FreeInference.

\begin{figure}[t]
    \vspace{0mm}
    \centering
    \includegraphics[width=\linewidth]{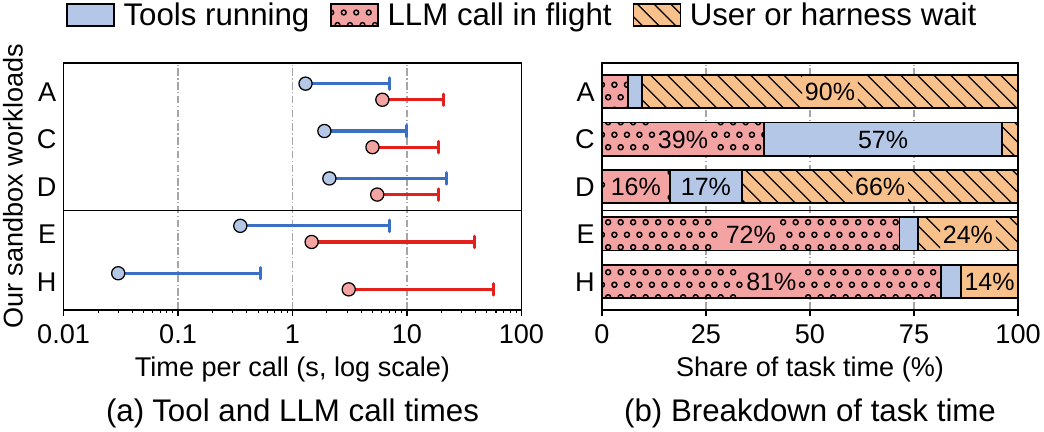}
    \vspace{-5mm}
    \caption{Tool and LLM call times of workloads A--H (median, line to P90) and breakdown of task time.}
    \label{fig:prod-tool-llm}
    \vspace{-1mm}
\end{figure}

Workloads A--H show the same trend.
Figure~\ref{fig:prod-tool-llm}a shows the median and
P90 tool and LLM call times of each workload. The
median tool call takes between 0.03\,s and 2.1\,s, while
the median LLM call takes between 1.5\,s and 6.1\,s.
Figure~\ref{fig:prod-tool-llm}b breaks down the
task time. Tools run for less than 5\% of the task time in A, E,
and H, and for 17\% in D. Interactive sandboxes spend most of their time waiting for the user: 90\% in A and 66\% in D. Sandboxes that run RL and evaluation jobs spend most of their time waiting for the LLM: 72\% in E and 81\% in H.
Both the open-source sandboxes and sandboxes A--H 
execute tools in bursts and then remain idle for seconds
during LLM calls and for minutes to hours between user turns.

\begin{figure}[t]
    \vspace{0mm}
    \centering
    \includegraphics[width=\linewidth]{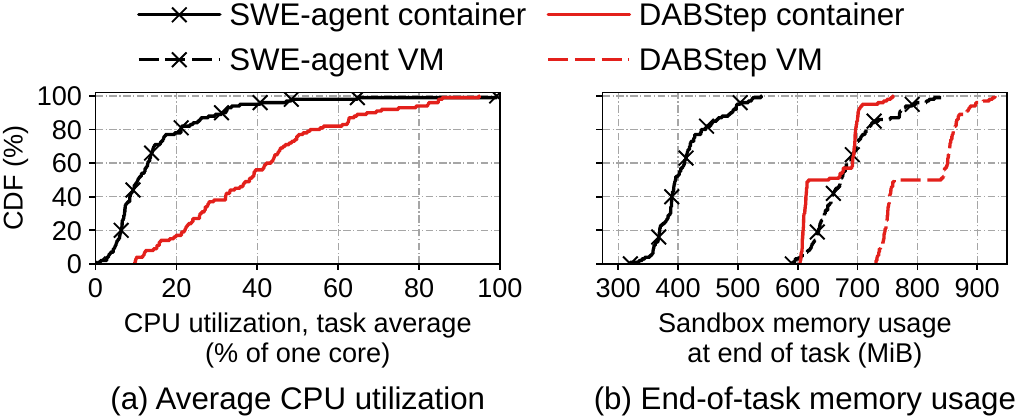}
    \vspace{-5mm}
    \caption{CPU utilization and memory usage distributions.}
    \label{fig:cpu-mem-dist}
    \vspace{-2mm}
\end{figure}

\vspace{2pt}
\noindent \textbf{CPU and memory utilization.}
In agentic workloads, CPU utilization spikes when a tool executes and drops to nearly
zero while the agent waits for the LLM. In contrast, memory
usage grows as tools read files and start processes, and stays
high for the rest of the task. Figure~\ref{fig:cpu-mem-dist}
shows the distributions of the average CPU utilization of each
task and of the sandbox memory usage at the end of each task.
CPU utilization is low: half of the tasks use $<$10.5\% and
38\% of a core in SWE-agent and DABStep, respectively. However,
at the end of the task, the median container holds 396\,MB
in SWE-agent and 635\,MB in DABStep. About half of this
memory is the page cache of the files that tools accessed, and
the rest is process memory.

\begin{figure}[t]
    \vspace{0mm}
    \centering
    \includegraphics[width=\linewidth]{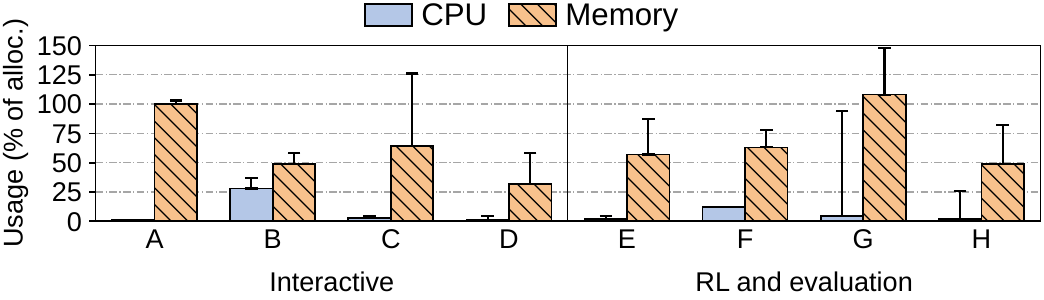}
    \vspace{-5mm}
    \caption{CPU and memory usage of workloads A--H as a share of the sandbox allocation (median, whisker to P90).}
    \label{fig:prod-cpu-mem}
    \vspace{-2mm}
\end{figure}

Figure~\ref{fig:prod-cpu-mem} shows the CPU and memory usage of
sandboxes A--H relative to their allocation. In a typical
minute, the median sandbox uses only 0.9--4.6\% of its CPU
allocation. B and F, which run short-lived VMs, use a larger
fraction (28\% and 12\%), yet still far less than they reserve.
Memory usage is much higher: the median sandbox uses 32--108\% of
its memory allocation. VMs hold the most memory. For example,
the median VM in A uses its whole allocation.
Hence, the CPU and
memory demands of agentic sandboxes are mismatched: sandboxes
reserve CPUs that they rarely use, while they occupy memory for
their whole lifetime.

\begin{figure}[t]
    \vspace{-1mm}
    \centering
    \includegraphics[width=\linewidth]{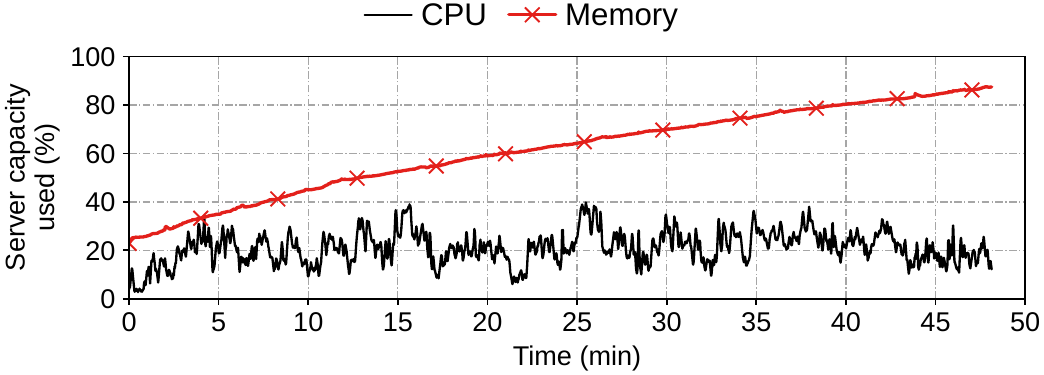}
    \vspace{-5mm}
    \caption{CPU utilization and memory usage when scaling agent density in the server.}
    \label{fig:agent-scaling}
    \vspace{-2mm}
\end{figure}

\begin{figure}[t]
    \vspace{0mm}
    \centering
    \includegraphics[width=\linewidth]{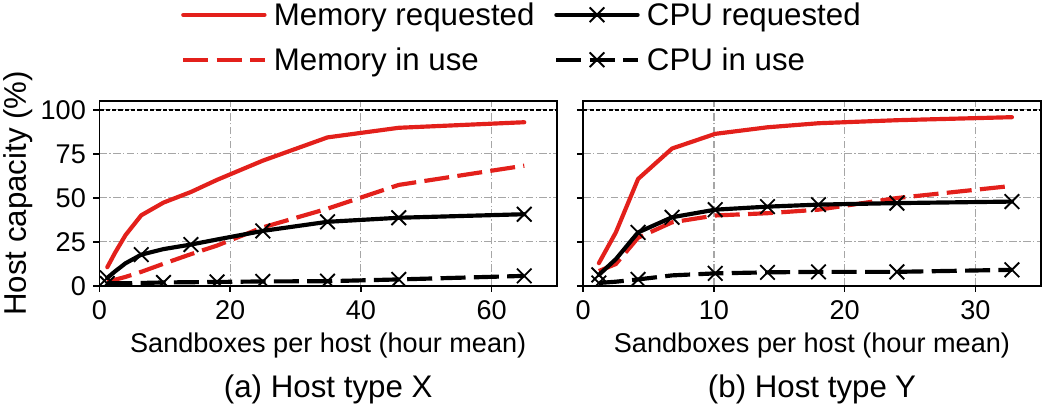}
    \vspace{-5mm}
    \caption{Host memory and CPU, requested and in use, as hosts pack more sandboxes.}
    \label{fig:prod-host-packing}
    \vspace{-4mm}
\end{figure}

\vspace{2pt}
\noindent \textbf{Scaling the number of concurrent agents.}
Figure~\ref{fig:agent-scaling} shows the CPU utilization and
memory usage of our server as we launch more concurrent SWE agents.
We set each agent to continuously run the same task back-to-back, i.e., without user-wait time.
We can see that memory utilization grows with each new agent and reaches 87.5\% of the
server's memory.
At that point, CPU utilization
is only 18\%. Hence, the server cannot admit more agents, even
though most of its cores are idle.
Figure~\ref{fig:prod-host-packing} shows
the memory and CPU that
our sandboxes request and use, as a fraction of the host's capacity,
as hosts pack more sandboxes. We show two host types. 
In both host types, requested memory grows quickly with the number of
sandboxes. On the most densely packed hosts, sandboxes request
93\% and 96\% of the host memory, and use 68\% and 57\%,
respectively. In contrast, they request 41\% and 48\% of the
cores, and use only 5.6\% and 9.1\%. Overall,
we can see that memory capacity limits the number of sandboxes per host.

\subsection{Memory Tiering}
\label{sec:char-tiering}

Memory tiering can free local DRAM for additional agents by
moving the memory of idle sandboxes to a slower tier. However,
it pays off only if the memory moves within the idle period and
the next tool invocation does not slow down.

\begin{figure}[t]
    \vspace{0mm}
    \centering
    \includegraphics[width=\linewidth]{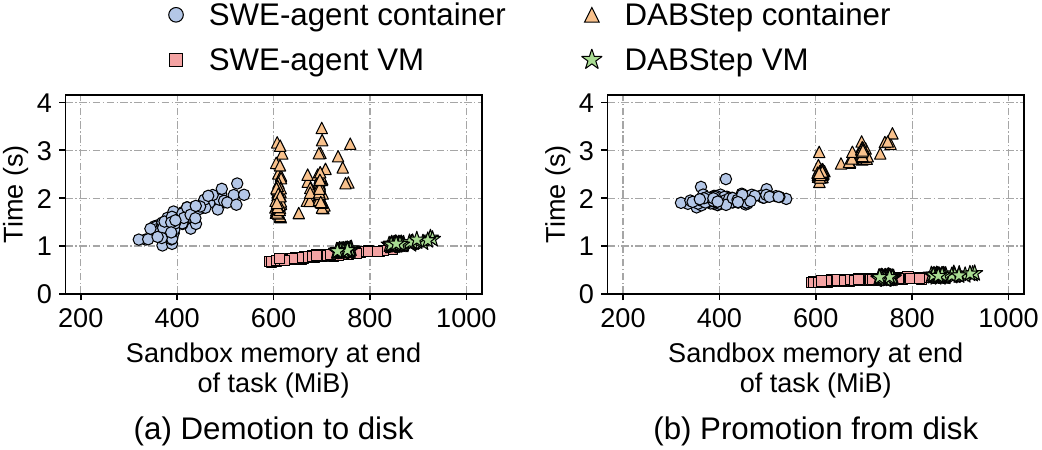}
    \vspace{-5mm}
    \caption{Disk offload times for containers and VMs.}
    \label{fig:disk-offload}
    \vspace{-2mm}
\end{figure}

\begin{figure}[t]
    \vspace{0mm}
    \centering
    \includegraphics[width=\linewidth]{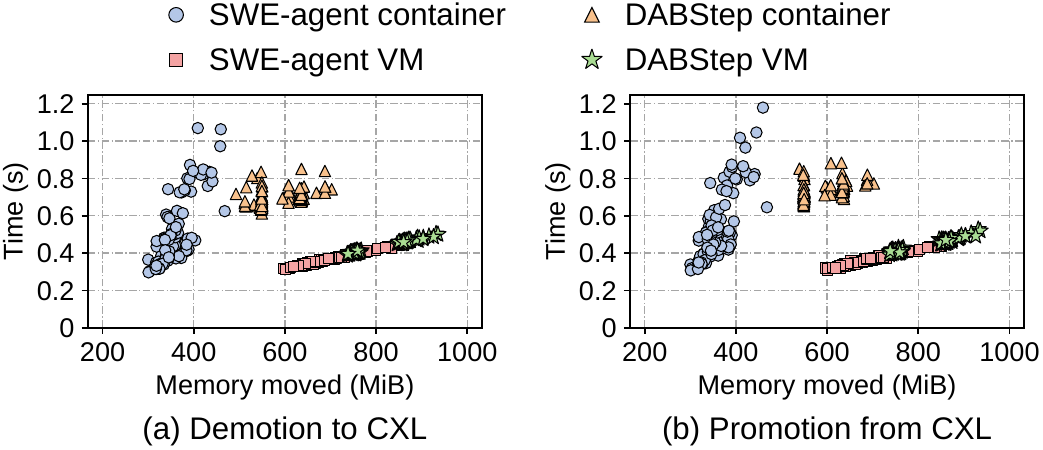}
    \vspace{-5mm}
    \caption{CXL offload times for containers and VMs.}
    \label{fig:cxl-offload}
    \vspace{-3mm}
\end{figure}

\vspace{2pt}
\noindent \textbf{Offload times to disk and CXL.}
To understand whether tiering fits within the sandboxes' idle periods, we demote the memory of each
sandbox at its peak memory and then promote it back. For
disk, a container's process memory goes to swap, and the page
cache of its files goes to a snapshot file. A VM's guest memory
is a single region, so we write its resident pages to a snapshot
file sequentially, and read the file back in one sequential
read. For CXL, we migrate the sandbox's pages between the DRAM
and the CXL.

Figure~\ref{fig:disk-offload} shows the time to demote a sandbox
to disk and to promote it back, as a function of the sandbox's
memory size. Each point is one task. At the median, containers take
1.5--2.1\,s to demote and 2.0--2.7\,s to promote. VMs are faster
even though they hold more memory: the median VM demotes in
0.77--0.97\,s and promotes in 0.29--0.36\,s. The difference
comes from the access pattern. A container's memory
returns from swap one page at a time from scattered locations,
at about 200--240\,MB/s. In contrast, a VM's memory returns in
one sequential read, at about 2.2--2.3\,GB/s. Hence, a disk
round trip takes 3.5--4.9\,s for a container, which is
comparable to a median LLM call, but only 1.1--1.3\,s for a VM.

Figure~\ref{fig:cxl-offload} shows the same measurements for CXL
memory. Migration to and from CXL is symmetric and faster than
disk: at the median, a container moves in 0.47--0.73\,s in each
direction, and a VM in 0.36--0.44\,s. Again, VMs move memory
faster, at about 1.9\,GB/s against 0.7--0.8\,GB/s for
containers.
Overall, a CXL round trip takes 0.7--1.4\,s, which fits within a
median LLM call.

In both tiers, the time grows with the memory moved.
For a given mechanism and sandbox type, the rates of SWE-agent
and DABStep differ by $<$10\%, with one exception:
containers on disk, whose rates differ by 20\%. The reason
is how a container's memory is split between process
(anonymous) memory and page cache. When we promote a container
from disk, its process memory returns from swap at
100\,MB/s, while its page cache returns from the snapshot
file at close to 1\,GB/s.
Thus, process memory takes
92--94\% of the promotion time, although it is only 45--53\% of
the memory moved, and DABStep, which has a larger share of page
cache, moves faster. VMs do not show this effect, as the guest's
page cache is part of guest memory. A memory manager can
estimate the time to demote and promote a sandbox from its size,
its type, and the split of its memory, and select a mechanism
that fits the expected idle period.

\begin{figure}[t]
    \vspace{0mm}
    \centering
    \includegraphics[width=\linewidth]{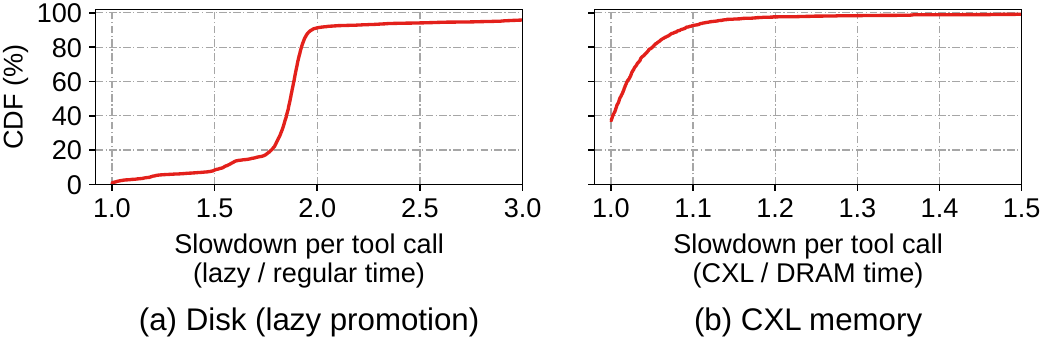}
    \vspace{-5mm}
    \caption{SWE-agent slowdown relative to DRAM with lazy promotion from disk and with execution from CXL memory.}
    \label{fig:tool-slowdown}
    \vspace{-2mm}
\end{figure}

\vspace{2pt}
\noindent \textbf{Tool slowdown with lazy promotion and remote memory access.}
A memory manager can promote pages lazily, on first
access, or leave them in CXL memory, which the CPU accesses
directly. Figure~\ref{fig:tool-slowdown}a shows the slowdown of
SWE-agent tool calls when their memory is promoted lazily from disk.
Since each first access to a page faults and reads the page from
swap, the median tool call runs 1.9$\times$ slower than from DRAM.
Figure~\ref{fig:tool-slowdown}b shows the slowdown of each tool
call when the sandbox runs from CXL memory. The median tool call
slows down by only 1\%. However, a few tool calls slow down by up to 1.5$\times$.
Hence, lazy promotion from disk puts page faults on the critical
path, while CXL memory slows down some tools. To avoid both
overheads, a memory manager needs to promote the pages that the
next tool needs before the tool starts.

\begin{figure}[t]
    \vspace{0mm}
    \centering
    \includegraphics[width=\linewidth]{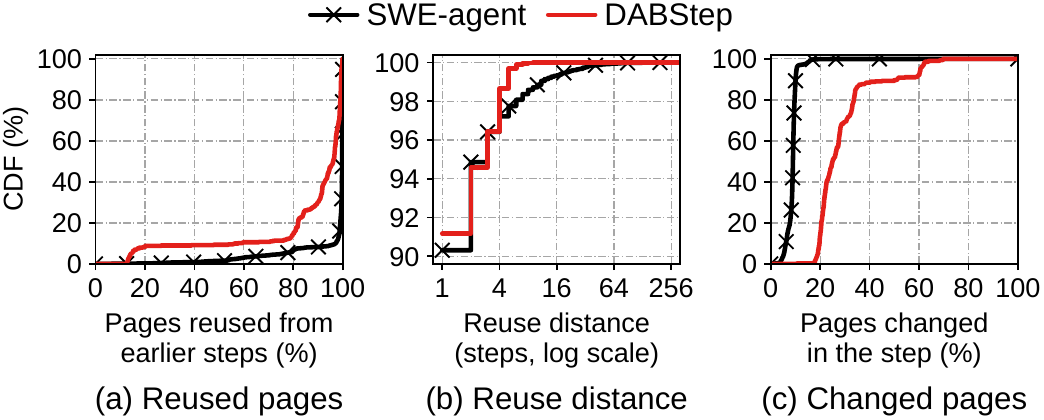}
    \vspace{-5mm}
    \caption{Fraction of each step's pages reused from earlier steps, reuse distance in steps, and fraction of dirty pages.}
    \label{fig:page-cdf}
    \vspace{-3mm}
\end{figure}

\vspace{2pt}
\noindent \textbf{Page reuse across steps.}
To learn whether the next tool's memory is
predictable, we record the pages that each step of a task
accesses and modifies.
Figure~\ref{fig:page-cdf}a shows the distribution of the
fraction of each step's pages that earlier steps already used.
Reuse is high: the median step reuses 99.8\% of its pages
in SWE-agent and 96.1\% in DABStep. Figure~\ref{fig:page-cdf}b
shows the reuse distance, i.e., the steps since the
page's previous use.
90\% of the reused pages were used in the previous step, and
more than 99\% within the last 16 steps.
Figure~\ref{fig:page-cdf}c shows that a step also modifies few
pages: the median step modifies 9\% of its pages in
SWE-agent and 25\% in DABStep.
Hence, the most recent steps predict the pages that the next tool
needs, and a snapshot only needs to save the modified pages.

\begin{figure}[t]
    \vspace{0mm}
    \centering
    \includegraphics[width=\linewidth]{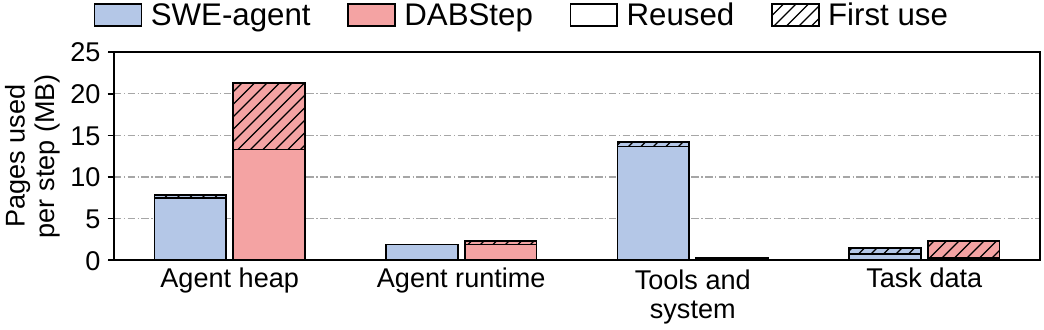}
    \vspace{-5mm}
    \caption{Pages used per step by source, split into pages reused from earlier steps and pages used for the first time.}
    \label{fig:page-types}
    \vspace{-2mm}
\end{figure}

\begin{figure}[t]
    \vspace{0mm}
    \centering
    \includegraphics[width=\linewidth]{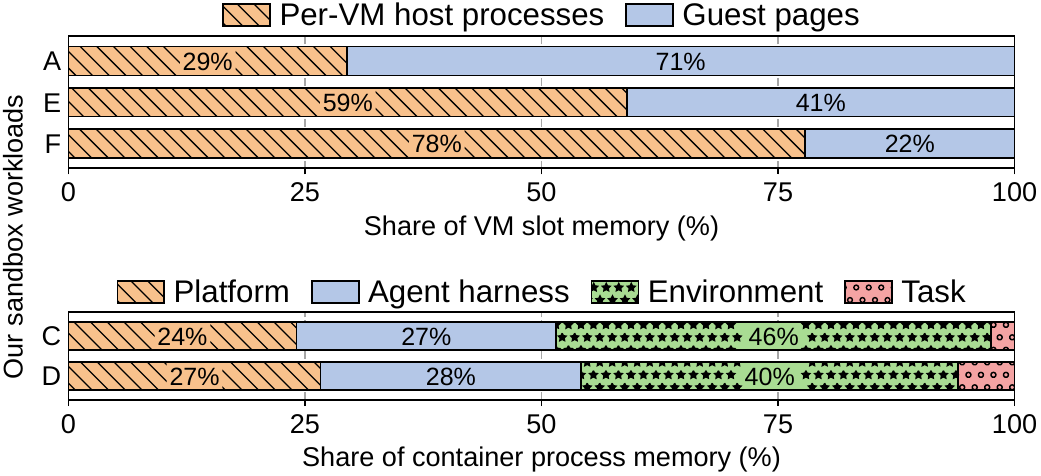}
    \vspace{-5mm}
    \caption{Sandbox memory breakdown: per-VM host and guest pages in VMs, and process roles in containers.}
    \label{fig:prod-memory-composition}
    \vspace{-4mm}
\end{figure}

\vspace{2pt}
\noindent \textbf{Page type breakdown.}
Figure~\ref{fig:page-types} breaks down the pages that a step
uses by their source: the agent's heap, the agent's runtime
(its interpreter and libraries), the tools and system software,
and the task's data. Each bar is split into pages reused from
earlier steps and pages used for the first time. On average, a
step uses 25\,MB in both agents, i.e., only
4--6\% of the sandbox memory. In SWE-agent, most of these pages
are tools and system software (56\%) and the agent
heap (31\%), which repeat in every step: 96\% of them are
reused. In DABStep, the agent heap dominates (81\%), as the agent
executes its Python code in its own interpreter, and 63\% of the
heap pages are reused. Task data is the only source dominated by
first uses: only 49\% and 11\% of its pages are reused in
SWE-agent and DABStep, respectively.
Hence, most of the
next tool's working set is known in advance, and only the task
data needs to be fetched on demand.

Figure~\ref{fig:prod-memory-composition} shows that, in
our sandbox workloads, the task also holds only a small
part of the memory. In VMs, the per-VM host processes that run
beside the guest take 29--78\% of the sandbox memory, and the
guest's pages take the rest. In containers, the task's own
processes hold only 2.5--6\% of the memory, while the
environment (40--46\%), the agent harness (28\%), and the
platform (24--27\%) hold the rest. These components stay
resident through every idle period, and they are largely the
same across steps. Hence, they offer a large and predictable
opportunity for memory tiering.

\subsection{Memory Compression}
\label{sec:char-compression}

Memory compression reduces the size of idle sandboxes, which could create space for additional agents. However, it
consumes CPU cycles and adds decompression latency when the sandbox resumes.
To understand how much memory is freed by compression, we compress the sandbox at its peak memory utilization. We run with compression algorithms commonly used in datacenters~\cite{karandikar2023cdpu,jeong2023compression}: \emph{zstd} and \emph{lz4}.
We run zstd with levels 1, 3, and 9. Higher levels achieve higher compression ratios but are more compute-heavy.

\begin{figure}[t]
    \vspace{0mm}
    \centering
    \includegraphics[width=\linewidth]{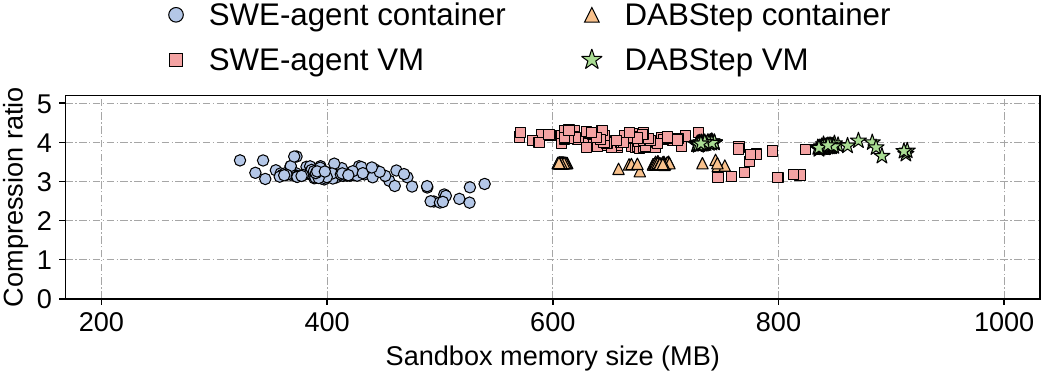}
    \vspace{-5mm}
    \caption{Compression ratios across workloads and sandbox types with the zstd-1 algorithm.}
    \label{fig:cdf-compression-ratio}
    \vspace{-3mm}
\end{figure}

\begin{figure}[t]
    \vspace{0mm}
    \centering
    \includegraphics[width=\linewidth]{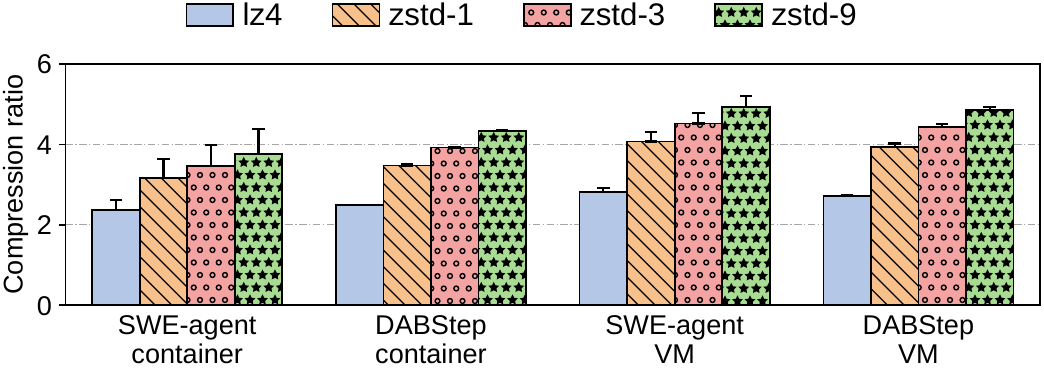}
    \vspace{-5mm}
    \caption{Median and P99 compression ratios of different algorithms across workloads and sandbox types.}
    \label{fig:bar-compression-ratio}
    \vspace{-4mm}
\end{figure}

\vspace{2pt}
\noindent \textbf{Compression ratios.}
Figure~\ref{fig:cdf-compression-ratio} shows the compression
ratio of sandboxes with the zstd-1 algorithm across workloads and sandbox types, as a function of the sandbox's
memory size. The median ratio is 3.2 and 3.5 for SWE-agent and
DABStep containers, and 4.1 and 3.9 for the VMs. Hence,
compression frees 69--76\% of the sandbox memory.
Figure~\ref{fig:bar-compression-ratio} shows the compression ratio with different algorithms across workloads and sandbox types.
With lz4, the ratio is 25--31\% lower than with zstd-1, while
zstd-3 and zstd-9 raise it by 9--13\% and 19--25\%.
This shows that stronger algorithms free more memory, but, as we show
next, they take longer to compress.

\begin{figure}[t]
    \vspace{0mm}
    \centering
    \includegraphics[width=\linewidth]{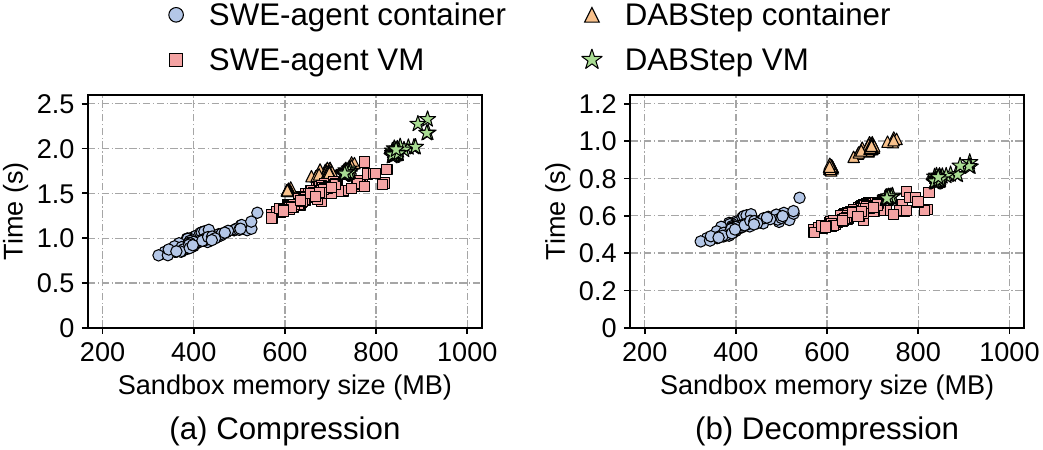}
    \vspace{-5mm}
    \caption{Compression and decompression times across workloads and sandbox types with the zstd-1 algorithm.}
    \label{fig:cdf-compression-time}
    \vspace{-2mm}
\end{figure}

\begin{figure}[t]
    \vspace{0mm}
    \centering
    \includegraphics[width=\linewidth]{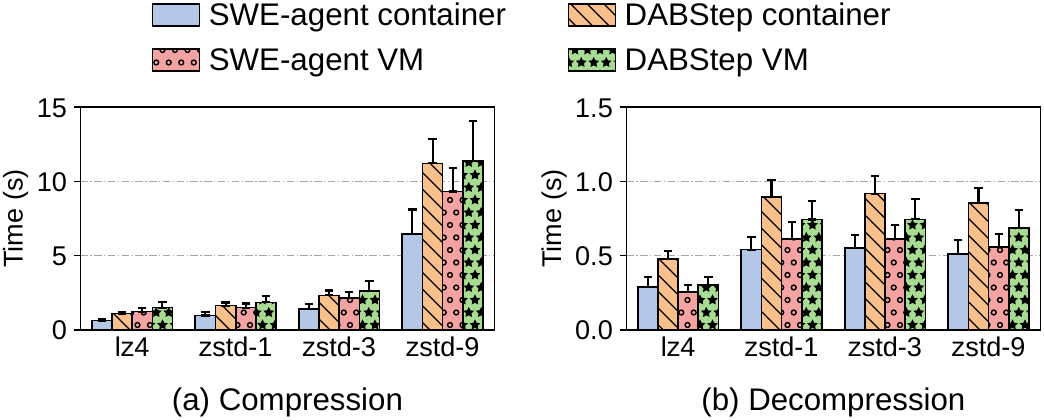}
    \vspace{-6mm}
    \caption{Median and P99 (de)compression times of different algorithms across workloads and sandbox types.}
    \label{fig:bar-compression-time}
    \vspace{-1mm}
\end{figure}

\vspace{2pt}
\noindent \textbf{(De)compression times.}
Figure~\ref{fig:cdf-compression-time} shows the time to compress
and decompress the memory of sandboxes with the zstd-1 algorithm across workloads and sandbox types. Both
times grow linearly with the memory size. At the median,
compression takes 0.96--1.85\,s, at 400--450\,MB/s, and
decompression takes 0.54--0.90\,s, at 0.7--1.1\,GB/s.
Hence, a
compression round trip takes 1.5--2.6\,s, which is slower than a
CXL round trip but still fits within a median LLM call. However, the
CPU executes all of this work, so compression consumes the cores
that the additional agents need.
Figure~\ref{fig:bar-compression-time} shows the time to compress and decompress
the memory across workloads and sandbox types with different algorithms.
Compared to zstd-1, lz4 compresses in 18--35\% less time and
decompresses 1.9--2.5$\times$ faster. In contrast, zstd-9
compresses 6.2--6.9$\times$ slower, which is
longer than the median LLM call of every workload we study.
Decompression time barely changes across zstd levels.
Overall, no
single algorithm fits all idle periods: a fast algorithm fits
LLM calls, while a strong one fits the long waits for the user.

\begin{figure}[t]
    \vspace{0mm}
    \centering
    \includegraphics[width=\linewidth]{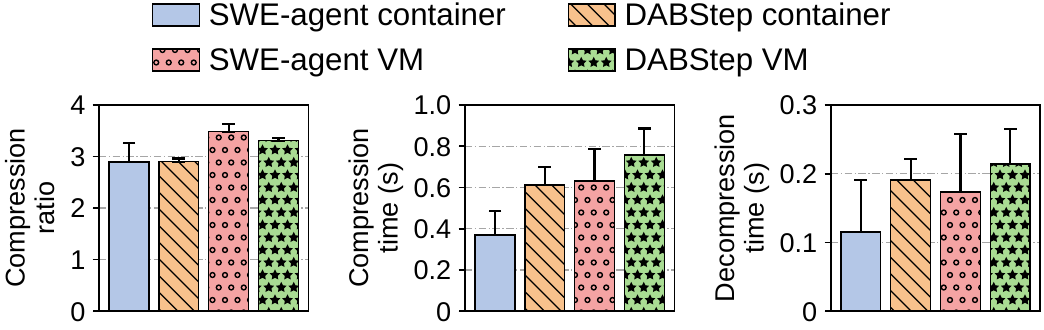}
    \vspace{-5mm}
    \caption{Compression ratio and (de)compression times with the IAA accelerator across workloads and sandbox types.}
    \label{fig:compression-change}
    \vspace{-4mm}
\end{figure}

\vspace{2pt}
\noindent \textbf{Opportunities with hardware (de)compression accelerators.}
Figure~\ref{fig:compression-change} shows the compression ratio and (de)compression times
when using
the IAA accelerator on our Intel server.
IAA compresses a
sandbox in 0.37--0.76\,s and decompresses it in 0.12--0.21\,s.
Compared to zstd-1 in software, compression is 2.4--2.7$\times$
faster and decompression 3.5--4.7$\times$ faster. The ratio is
9--17\% lower than with zstd-1 (2.9--3.5), but still 17--24\%
higher than with lz4. Hence, an IAA round trip takes
0.5--1.0\,s, which is comparable to a CXL round trip, and the
cores remain free for agents. However, all sandboxes on a socket
share only a few accelerator engines (e.g., 8--24), so a memory manager needs to coordinate
its use across sandboxes.

\section{\system{}: Efficient Memory Management for Agentic AI Sandboxes}
\label{sec:design}

\begin{figure}[t]
    \vspace{0mm}
    \centering
    \includegraphics[width=\linewidth]{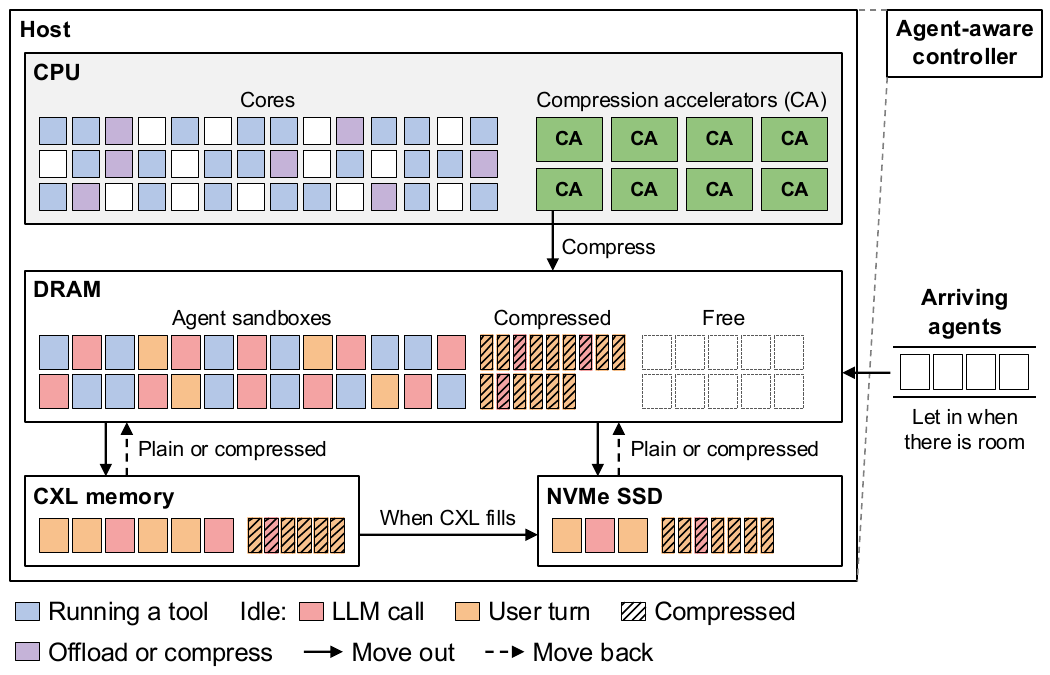}
    \vspace{-6mm}
    \caption{High-level architecture of \system{}.}
    \label{fig:overview}
    \vspace{-6mm}
\end{figure}

Based on our insights,
we propose \system{}, an agent-aware memory manager that
increases the effective memory capacity of a server. \system{}
reclaims the memory of idle sandboxes when the server needs
more space, and it restores that memory when the agent needs its
state again. Figure~\ref{fig:overview} shows a host with
\system{}. Sandboxes execute tools on CPU cores. Their memory
is in local DRAM, CXL memory, or an NVMe SSD, either plain
or compressed. Compression runs either in software on the cores
or on a hardware accelerator, such as Intel IAA~\cite{inteliaa}
or QAT~\cite{intelqat}. \system{}'s agent-aware controller tracks
the state of each sandbox: running a tool, waiting for an LLM
call, or waiting for the user. It uses this state to manage
memory as agents arrive and idle sandboxes resume. When an
agent arrives and local DRAM has enough free space, the
controller admits the agent directly.
Otherwise, it tries to free space by moving or compressing the memory of
idle sandboxes, and admits the agent only if it frees enough.
Later, before an idle sandbox resumes, the controller restores
its memory to local DRAM using the same procedure.

\system{} has two components. First, a decision policy
determines how to make enough space for new and returning
agents (\S~\ref{sec:design-placement}). For each idle
sandbox, it weighs the local memory that each operation would
free against the time that the operation needs to reclaim and
restore the memory, and against the estimated time until the sandbox needs
its memory again. Second, three optimizations reduce the
overhead of these operations (\S~\ref{sec:design-opt}): 
(i) efficient snapshotting writes a container's scattered memory
sequentially and moves only dirty pages, (ii) hot-page tracking
restores only the pages that the next tool needs, and (iii) hardware
accelerators perform data movement and compression.

\vspace{-2mm}
\subsection{\system{} Reclamation and Restoration Policy}
\label{sec:design-placement}

\system{} uses the insight that the state of an agent reveals
when its sandbox will need its memory again. A sandbox that runs
a tool needs its memory now, one that waits for an LLM call
needs it within seconds, and one that waits for the user needs
it within minutes or hours. Conventional memory managers, such
as TMO~\cite{weiner2022tmo} and TPP~\cite{maruf2023tpp}, select
individual pages based on recent accesses.
\system{} manages memory per
sandbox, and uses the agent's state to determine how much time
it has to reclaim and restore each sandbox's memory. To make deployment practical, the policy is deliberately simple. It
visits idle sandboxes from the longest to the shortest predicted
idle time, and places each one in the first memory tier that it
can return from in time. It needs no solver and no per-page
profiling, and it relies only on 
signals available from the agent harness, host, and inference interface.

\vspace{2pt}
\noindent \textbf{Reclamation policy.}
\system{} admits arriving agents if local DRAM usage
is below a high watermark. When a new agent exceeds
 it, the controller reclaims memory from idle agents
until the usage drops below a low watermark. Sandboxes that run
a tool are never reclaimed. The controller visits the idle
agents in decreasing order of predicted remaining idle time,
so that it first reclaims the memory that will be needed last.
Hence, it visits those that wait for the user before
sandboxes that wait for an LLM call. For each visited sandbox,
the controller selects a placement that \emph{fits}: the target
has room, and the predicted time to reclaim and restore the
sandbox, plus a safety margin, is shorter than its predicted
remaining idle time.

A sandbox that waits for the user moves to the SSD. Its idle
period hides even the slowest round trip, and it leaves the
faster CXL memory to sandboxes that return within seconds. A
sandbox that waits for an LLM call tries three placements in
order. First, it moves to CXL memory, which frees all of its
DRAM without any compression work. If CXL memory is short of
space, it moves to CXL memory in compressed form, which takes
about a third of the space. If neither placement fits, it is
compressed in local DRAM, which frees about 70\% of its memory
with the shortest round trip. If no placement fits, the sandbox
stays in local DRAM. The controller stops as soon as it frees
enough memory. If it visits all idle sandboxes without freeing
enough memory, it rejects the arriving agent, and the cluster
scheduler places the agent on another host. The same policy
manages CXL memory. When CXL usage crosses its own high
watermark, the controller moves the sandboxes in CXL memory
that wait for the user to the SSD.

\vspace{2pt}
\noindent \textbf{Restoration policy.}
\system{} starts to promote a sandbox ahead of its predicted
next tool, by the predicted promotion time, so that its memory
is back in local DRAM when the tool starts. If local DRAM has
no room, the controller frees it with the same rules. A
returning sandbox has priority over a new arriving agent. Hence,
the space between the high watermark and the DRAM capacity is
reserved for returning sandboxes, and the controller rejects
arriving agents before it delays a returning sandbox. There are
three exceptions. First, if a sandbox resides in CXL memory and
its next tool is insensitive to remote memory accesses, the
sandbox stays in CXL memory and the tool runs on the local
cores. \system{} records the slowdown of each tool when it runs
from CXL memory, and treats a tool as insensitive if that
slowdown is below a threshold (e.g., 5\%). Second, if a tool
starts before its sandbox is fully promoted, the remaining
pages are promoted on first access. Therefore, a misprediction
slows the tool down but never stalls it. Third, if a sandbox
becomes active while it is being demoted, the controller
cancels the demotion and promotes the pages that already moved.

\vspace{2pt}
\noindent \textbf{Predicting overheads and idle times.}
The policy needs two predictions. First, it needs the time to reclaim and restore a sandbox. As \S~\ref{sec:char-tiering} shows, this time depends on the mechanism, the sandbox type, the sandbox memory, and its split between process memory and page cache, and it grows linearly with the memory moved. Hence, \system{} profiles each mechanism offline and fits a linear regression model per mechanism and sandbox type, with the sandbox memory and its split as inputs. \system{} then refines each prediction with the history of the same task: the times that it measured when it moved the sandbox in prior steps and turns. The history also gives the compression ratio of the sandbox, which determines how much memory compression frees.

Second, the policy needs the remaining idle time of a sandbox, which \system{} predicts from the cause of the wait. For an LLM call, \system{} 
can use any of the LLM output length predictors proposed by prior work~\cite{xie2026entropy,zheng2023response,qiu2024proxy,stojkovic2025dynamollm}.
In our prototype, 
\system{} 
relies on 
the hidden states of the serving LLM~\cite{xie2026entropy}. A lightweight head pools the hidden states of the prompt tokens, weighted by their entropy, and predicts the expected output length.  
Then, analytical models estimate the time to first token (TTFT) from the input length and the time between tokens (TBT) from the load of the inference server. The predicted wait is the TTFT plus the expected output length times the TBT. For a user turn, \system{} uses the idle times between the prior turns of the same session, and falls back to the distribution of idle times across sessions~\cite{liu2026agenticcoding} for the first turns. It conditions the prediction on the time that has already elapsed, since a user who has been idle for long is likely to stay idle. In both cases, \system{} uses a conservative estimate, so that errors favor promoting early.

Finally, \system{} corrects its predictions with two early signals of the next tool. The first signal is an \emph{LLM hint}. When the inference server decodes the start of a tool call, it notifies the controller before the rest of the response is generated and returned. The second signal is the arrival of the user's next turn. Before the next tool can run, an LLM call must process the turn. On either signal, the controller starts to promote the sandbox if it has not started yet. Hence, the rest of the LLM call hides part of the promotion.

\subsection{Reducing the Overhead of Sandbox Memory Reclamation and Restoration}
\label{sec:design-opt}

The faster \system{} reclaims and restores memory,
the more sandboxes fit in the slower tiers, and the more local
DRAM it frees. 
Thus, \system{}
introduces three optimizations to minimize these overheads.

\vspace{2pt}
\noindent \textbf{Efficient snapshotting.}
\S~\ref{sec:char-tiering} shows that containers move their
memory slower than VMs. A VM's guest memory is a
single region, which \system{} writes and reads sequentially. In
contrast, a container's memory is scattered across the address
spaces of its processes and the page cache of its files. To
reclaim it, the kernel must find each page through several
levels of indirection and write it to a scattered swap slot. To
restore it, the kernel reads the pages back one at a time.
Hence, process memory takes 91--94\% of the promotion time of a
container, although it is less than half of its memory.
Further, every reclamation writes the whole sandbox again, even
though most of its pages have not changed.

To address these issues, \system{} keeps a \emph{snapshot table}
for each sandbox. Each entry records a resident page, its owner
(a process and virtual address, or a file and offset), its
offset in the snapshot, and whether it is dirty. \system{} builds
the table at the first reclamation of the sandbox, and then
maintains it incrementally. With the table, \system{}
\emph{gathers} the scattered pages of a container and writes
them to one contiguous snapshot in the target tier. On
restoration, it reads the snapshot sequentially and
\emph{scatters} each page back to its owner, without searching
for it. Hence, a container moves its memory with the same
sequential access pattern as a VM. In addition, \system{} keeps
the snapshot in the slower tier after it restores the sandbox,
and tracks the pages that the sandbox writes while it runs. It
uses soft-dirty bits for containers and the KVM dirty log for
VMs. At the next reclamation, \system{} drops the clean pages
from local DRAM without writing them, since the snapshot already
holds them. It writes only the dirty and new pages, as a
\emph{delta} to the snapshot. 
Hence, after the first reclamation, each reclamation writes at most a few percent of the sandbox. When a slower tier runs short of space, \system{} discards the snapshots of running sandboxes first, and writes their next reclamation in full.

\begin{figure}[t]
    \vspace{0mm}
    \centering
    \includegraphics[width=\linewidth]{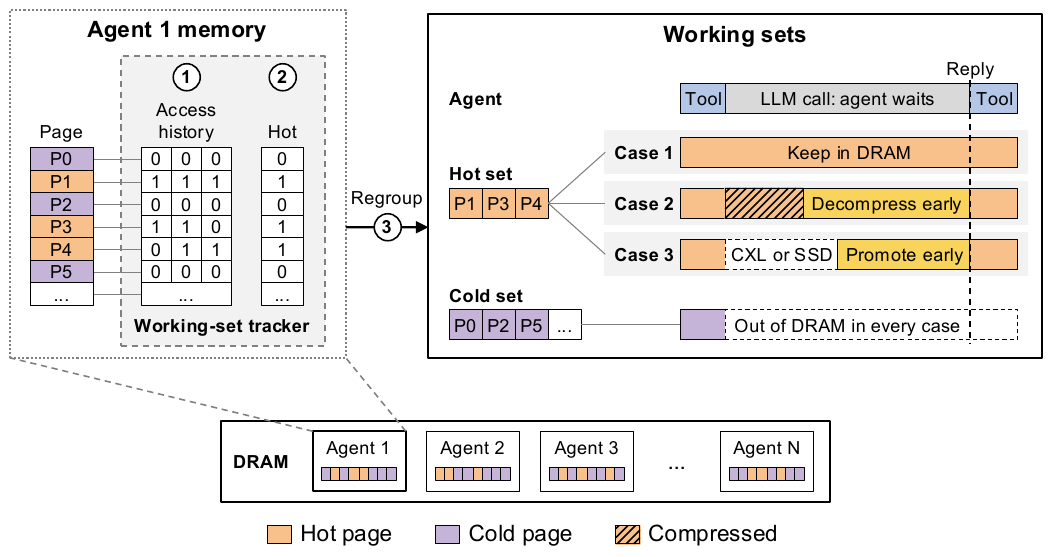}
    \vspace{-6mm}
    \caption{Hot set tracker overview.}
    \label{fig:hot-set-tracker}
    \vspace{-6mm}
\end{figure}

\vspace{2pt}
\noindent \textbf{Hot-page tracking.}
A full restoration moves the whole sandbox, although the next
tool accesses only a small part of it. In contrast, lazy
restoration puts a page fault on the critical path of every
first access.
\system{} avoids both overheads by restoring early only the
pages that the next tool is likely to access.
\S~\ref{sec:char-tiering} shows that these pages are
predictable: the median step reuses 96--99.8\% of its pages from
earlier steps, and 90\% of the reused pages were used in the
previous step.
\system{} uses this predictability to track the hot pages of
each sandbox, as Figure~\ref{fig:hot-set-tracker} shows. A
working-set tracker keeps the access history of each page over the last few steps (1). After each
tool, \system{} scans the page table of the sandbox, records the
accessed bit of each page, and clears it. The same scan also
collects the dirty bits for snapshotting. A page that was
accessed in any of the recent steps is \emph{hot} (2). At
reclamation, \system{} regroups the pages of the sandbox into a
hot set and a cold set, and stores the hot set contiguously in
its snapshot (3). Then, the restoration depends on the placement
of the sandbox. If the sandbox stays in local DRAM, nothing
moves (Case 1). If the sandbox is compressed in local DRAM,
\system{} decompresses the hot set before the next tool (Case 2).
If the sandbox resides in CXL memory or on the SSD, \system{}
promotes the hot set before the next tool (Case 3). In all cases,
the cold set stays out of local DRAM, and its pages are promoted
on first access. Hence, a restoration moves a few percent of the
sandbox instead of all of it.
Further, when the controller checks whether a placement fits
(\S~\ref{sec:design-placement}), it uses the time to restore
only the hot set. As a result, more sandboxes fit in the slower tiers.

\vspace{1pt}
\noindent \textbf{Hardware acceleration.}
(De)compression and page copies consume CPU cycles
that agents need. \system{} offloads this work to 
accelerators that are already present in modern servers. For
example, it (de)compresses  memory with
Intel's IAA~\cite{inteliaa}. On average,
IAA compresses 2.5$\times$ faster and decompresses 4.1$\times$
faster than software (\S~\ref{sec:char-compression}).
\system{} could also copy pages between local DRAM and CXL memory with a data movement accelerator, e.g., Intel DSA~\cite{liu2025dsa2lm}. 
However, a server has only a few
accelerator engines, and many sandboxes can reclaim and restore
memory at the same time, e.g., when several LLM calls complete
together. To avoid contention, the \system{} controller schedules
all operations on the accelerators. It gives restorations
priority over reclamations, since a late restoration delays the
next tool, and it orders restorations by the predicted start
time of the next tool. Further, when the controller selects a
tier for an idle sandbox (\S~\ref{sec:design-placement}), it
checks that the sandbox can return before its next tool. In this
check, it adds the current queueing delay of the accelerators to
the predicted time to reclaim and restore the sandbox. If the accelerators are busy
and a restoration would not complete before the next tool,
\system{} falls back to software and parallelizes the work across
the idle cores of the server. Hence, \system{} reclaims and
restores memory quickly, with few CPU cycles, and without
delaying the tools.
\section{Evaluation}
\label{sec:evaluation} 

\label{sec:methodology}

\vspace{2pt}
\noindent \textbf{Experimental setup.}
We run experiments on an AMD Turin server 
with 96 cores, 128\,GB of local DRAM, a 128\,GB CXL
memory, and a 2.7\,TB NVMe SSD. Local DRAM and CXL use DDR5 at 3200\,MT/s and DDR4 at 2400\,MT/s, respectively. For LLM serving, we use servers with 8 NVIDIA A100-80GB GPUs. 
Since AMD host lacks on-die compression and our Intel host lacks CXL, we emulate IAA using per-sandbox ratios and latencies profiled on Intel IAA (\S~\ref{sec:char-compression}): we reserve the compressed footprint in DRAM/CXL, issue matching cache-bypassing memory reads and writes (retaining raw pages in uncounted memory), and model 6-IAA (8-engine per accelerator) queueing delays and CPU submission cycles.  

\vspace{3pt}
\noindent \textbf{Agentic workloads.}
We evaluate two agentic workloads, as in 
\S~\ref{sec:characterization}: 
(i) mini-SWE-agent~\cite{yang2024sweagent}
solving tasks from SWE-bench Verified~\cite{jimenez2024swebench} and (ii) 
DABStep agent~\cite{egg2025dabstep}, a smolagents CodeAgent,
solving data-analysis tasks. 
Both agents use
Qwen3-235B-A22B-Instruct~\cite{qwen3}, served by
vLLM~\cite{vllm}.
Each agent runs in its own sandbox. 
To evaluate isolated sandboxes without cross-sandbox page sharing, we give each agent a private copy of its files. 
As a result, the memory
of each agent, including its page cache, is charged to its own
cgroup, as it would be in a separate VM.

To keep LLM behavior consistent, we use
trace replay. We first run each benchmark and record every LLM
call, with its reply and inference latency, and every tool call.
Then, during the evaluation, each agent replays its trace. It
receives the recorded LLM replies after the recorded inference
latencies, and executes the tools in its sandbox. LLM
replies and their timing are fixed across configurations, the CPU and memory behavior of the sandboxes is real.

\vspace{3pt}
\noindent \textbf{Agent arrivals.}
Agents arrive with inter-arrival times drawn from the GitHub
Copilot traces from Azure~\cite{liu2026agenticcoding} (in total, there are 965 agents). After an
agent finishes a task, it stays idle for a time drawn from the
same traces, which emulates a user who reviews the result before
issuing the next request. Then, the agent starts a new task, and
repeats this cycle until the end of the experiment. An
experiment lasts 2 hours. 
Each agent reserves 800\,MB of
memory, which covers the peak memory of an agent that we measure.
When a system reclaims the memory of an agent, the
agent releases its reservation. The freed memory goes first to
the agents that wait to be restored, in FIFO order, and then to
new agents, which wait in an admission queue.

\begin{figure}[t]
    \vspace{0mm}
    \centering
    \includegraphics[width=\linewidth]{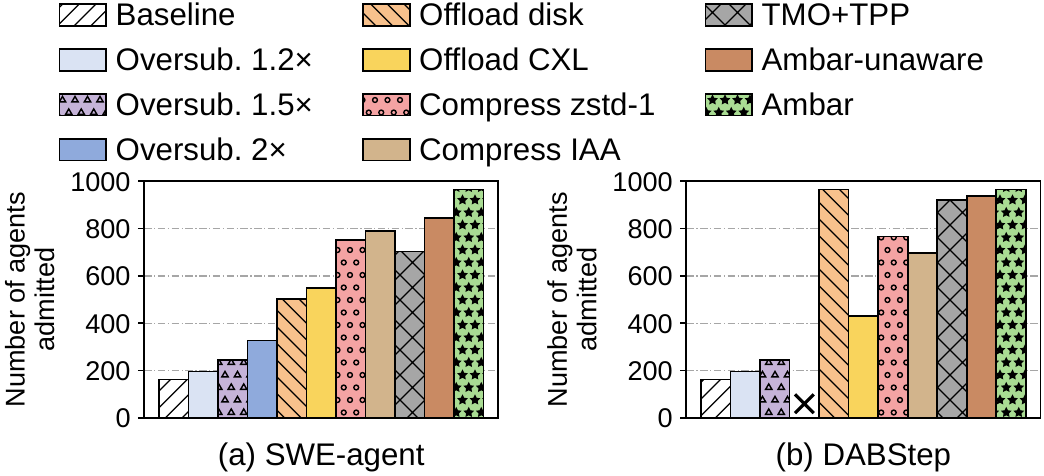}
    \vspace{-6mm}
    \caption{Number of agents admitted per server with each configuration across workloads.}
    \label{fig:agent-density}
    \vspace{-2mm}
\end{figure}

\vspace{3pt}
\noindent \textbf{Evaluated configurations.}
We 
evaluate eleven systems. 
\emph{Baseline} is conventional sandbox management, where each
agent holds its reservation for its whole lifetime.
\emph{Oversub.~$k\times$} admits agents against $k\times$ the
DRAM capacity, for $k$ = 1.2, 1.5, and 2.
Single-mechanism configurations reclaim the memory of an agent during LLM calls and idle periods. \emph{Offload disk} and \emph{Offload CXL}
move sandboxes to the SSD and to CXL memory, respectively.
\emph{Compress zstd-1} and \emph{Compress IAA} compress sandboxes with zstd-1 and with IAA,
respectively. These configurations restore an agent reactively,
when its wait ends.
\emph{TMO+TPP} is the Linux tiering stack: TMO-style proactive
reclaim~\cite{weiner2022tmo}, TPP demotion to
CXL~\cite{maruf2023tpp}, and zswap in front of SSD swap.
\emph{\system{}-unaware} uses all mechanisms of \system{}, but
places sandboxes by their idle time and restores them reactively. \system{} is our full system.

\subsection{Evaluation Results}

\vspace{2pt}
\noindent \textbf{1. Agent density.}
Figure~\ref{fig:agent-density} shows the number of agents that
each configuration admits in SWE-agent and DABStep. \emph{Baseline}
admits only 163 agents, as the reservations of the agents fill
local DRAM. The \emph{Oversub.} configurations admit more agents, but only up to
327 with \emph{Oversub.~2$\times$} in SWE-agent. In DABStep,
\emph{Oversub.~2$\times$} runs out of memory.
The single-mechanism configurations admit many more agents, as
they reclaim memory of idle sandboxes. However, each of them is
limited by its mechanism. For example, \emph{Offload CXL} admits only
430 agents in DABStep, since CXL memory fills up.
\emph{TMO+TPP} and \emph{\system{}-unaware} use several tiers, and
admit 703 and 844 agents in SWE-agent, respectively. However,
neither knows what an agent waits for, so both keep the sandboxes
that wait for the user in the fast tiers for too long. In contrast,
\system{} combines all tiers: it moves the sandboxes that wait
for the user to the SSD, and keeps the faster tiers for the
sandboxes that wait for an LLM call. Hence, \system{} admits the
most agents in both workloads: 
965 agents. 
Compared to \emph{Baseline}, \system{} admits 
5.9$\times$ more agents.
In SWE-agent, it admits 14\% more agents than
\emph{\system{}-unaware} and 22\% more than the best
single-mechanism configuration (\emph{Compress IAA}).
In DABStep, \emph{TMO+TPP} and \emph{\system{}-unaware} admit 921 and 939 agents.

\begin{figure}[t]
    \vspace{0mm}
    \centering
    \includegraphics[width=\linewidth]{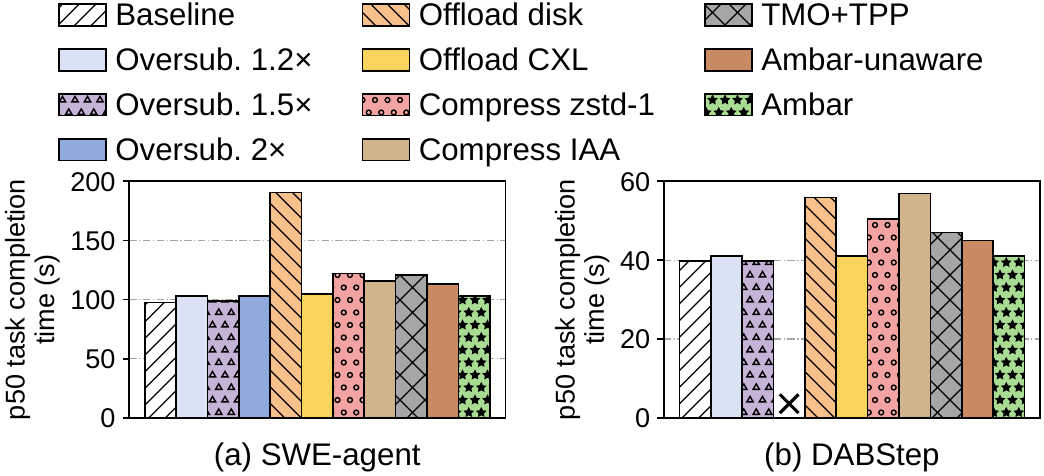}
    \vspace{-6mm}
    \caption{P50 task completion time with each configuration across workloads.}
    \label{fig:agent-latency-p50}
    \vspace{-3mm}
\end{figure}

\begin{figure}[t]
    \vspace{0mm}
    \centering
    \includegraphics[width=\linewidth]{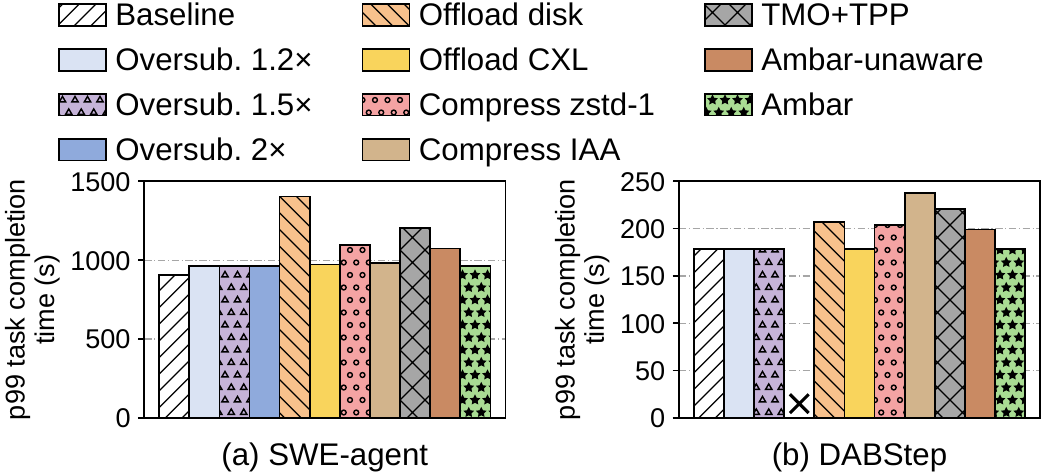}
    \vspace{-5mm}
    \caption{P99 task completion time with each configuration across workloads.}
    \label{fig:agent-latency-p99}
    \vspace{-2mm}
\end{figure}

\vspace{2pt}
\noindent \textbf{2. Agent latency.}
Figures~\ref{fig:agent-latency-p50} and~\ref{fig:agent-latency-p99}
show the P50 and P99 task completion times, respectively.
\emph{Baseline} reclaims no memory, and its P50 task takes 98\,s in
SWE-agent and 40\,s in DABStep. The single-mechanism
configurations put the restoration on the critical path of the
next tool, which increases the task time.
\emph{Offload disk} has the highest overhead: it increases the P50  by 95\% in SWE-agent and 41\% in DABStep, as the SSD
restores a sandbox slowly. \emph{Compress zstd-1} and \emph{Compress IAA}
increase the P50 task time by 18--43\%. \emph{Offload CXL} has a lower
overhead, but admits fewer agents. \emph{TMO+TPP} increases the P50
task time by 24\% in SWE-agent and 18\% in DABStep, as tools fault
their pages back from swap. \emph{\system{}-unaware} increases it by
16\% and 13\%, as it restores sandboxes reactively. In contrast, \system{} keeps
the task time close to \emph{Baseline}. It increases the P50 task time
by only 5.5\% in SWE-agent and 3.3\% in DABStep. It increases
the P99 task time by only 6.3\% in SWE-agent, and does not
increase it in DABStep, where \emph{TMO+TPP} increases it by
24\%. 
\system{} places each sandbox in a tier that it can return from
in time, and restores the hot pages before the next tool starts.

\begin{figure}[t]
    \vspace{0mm}
    \centering
    \includegraphics[width=\linewidth]{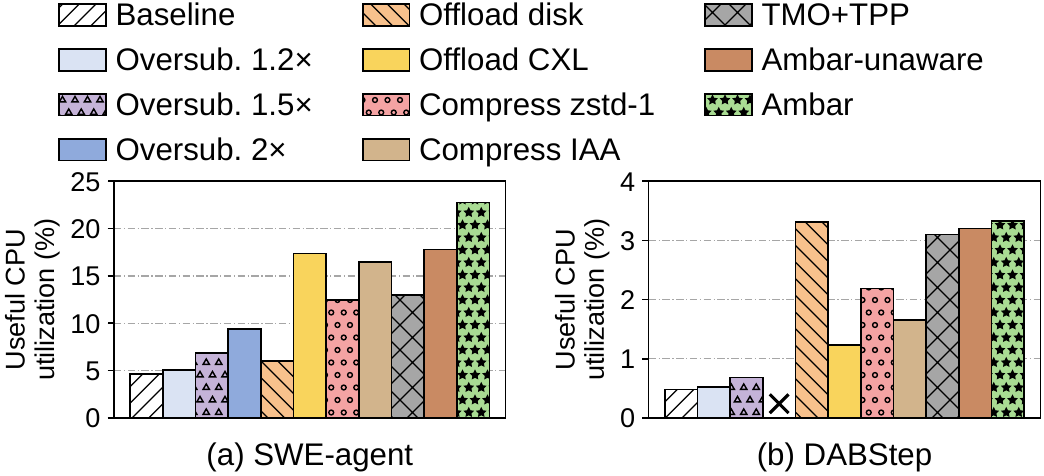}
    \vspace{-5mm}
    \caption{Useful CPU usage, without memory-management overhead, with each configuration across workloads.}
    \label{fig:cpu-util}
    \vspace{-3mm}
\end{figure}

\vspace{2pt}
\noindent \textbf{3. CPU utilization.}
Figure~\ref{fig:cpu-util} shows the useful CPU utilization of the
server, i.e., the CPU time that agents spend on their tools,
excluding the overhead of memory management. \emph{Baseline} uses only
4.7\% of the CPU in SWE-agent and 0.5\% in DABStep, as memory
limits the number of agents. \emph{Oversub.} increases the
utilization only slightly. The single-mechanism configurations
increase it further, but their overheads slow agents down.
For example, in SWE-agent, \emph{Compress zstd-1} admits nearly as many
agents as \emph{Compress IAA}, but its useful utilization is 24\% lower,
due to software compression. 
\system{} reaches 22.8\% in SWE-agent and 3.3\% in DABStep, which
is 4.9$\times$ and 6.6$\times$ higher than \emph{Baseline}, respectively.
In SWE-agent, \system{} achieves 28\% and 75\% higher useful
utilization than \emph{\system{}-unaware} and \emph{TMO+TPP}, respectively. In
DABStep, \emph{Offload disk} reaches the same utilization as \system{},
but with a 36\% longer P50 task time. Hence, \system{} packs more agents, while keeping their
tasks fast.

Utilization in DABStep is low in all
configurations, as its median tool call takes 1.0\,ms, while its
median LLM call takes 5.8\,s (\S~\ref{sec:char-workload}).
Hence, a server needs many agents, and thus much memory, to keep
its cores busy. With a higher load, \system{} increases the
utilization further (\S~\ref{sec:sensstudies}). For such
workloads, next-generation servers may need a higher ratio of
memory capacity to cores.

\subsection{Characterizing \system{} Operations}
\label{sec:charambar}

\begin{figure}[t]
    \vspace{0mm}
    \centering
    \includegraphics[width=\linewidth]{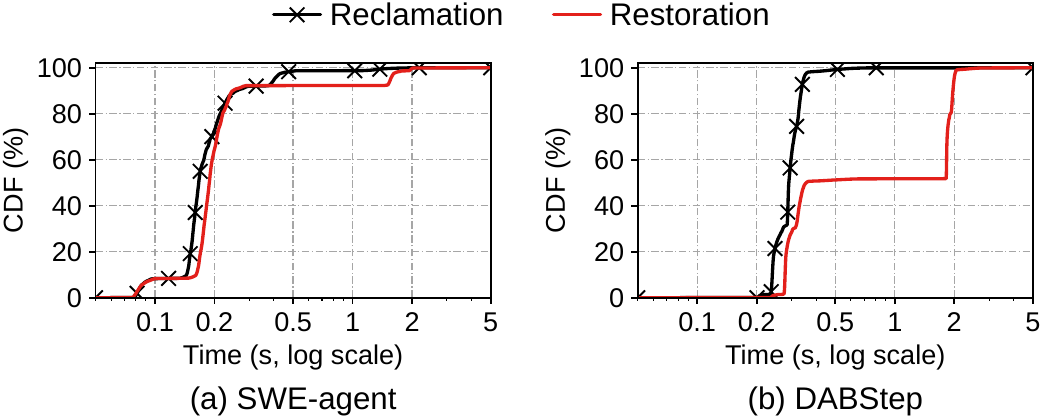}
    \vspace{-5mm}
    \caption{Distributions of the reclamation and restoration times of a sandbox with \system{} across workloads.}
    \label{fig:operation-times}
    \vspace{-2mm}
\end{figure}

\vspace{2pt}
\noindent \textbf{1. Distribution of memory reclamation or restoration times.}
Figure~\ref{fig:operation-times} shows the distributions of the
time that \system{} takes to reclaim and to restore the memory of
a sandbox. Most operations are short. In SWE-agent, the median
reclamation and restoration take 0.17\,s and 0.19\,s,
respectively.
In DABStep, the median reclamation takes 0.29\,s. However, the
restoration times in DABStep are bimodal: half of the
restorations take about 0.3\,s, and the other half take about
2\,s. The slow restorations bring sandboxes back from the SSD.
\system{} places only the sandboxes that wait for the user on the
SSD, as their idle periods last minutes.

\begin{figure}[t]
    \vspace{0mm}
    \centering
    \includegraphics[width=\linewidth]{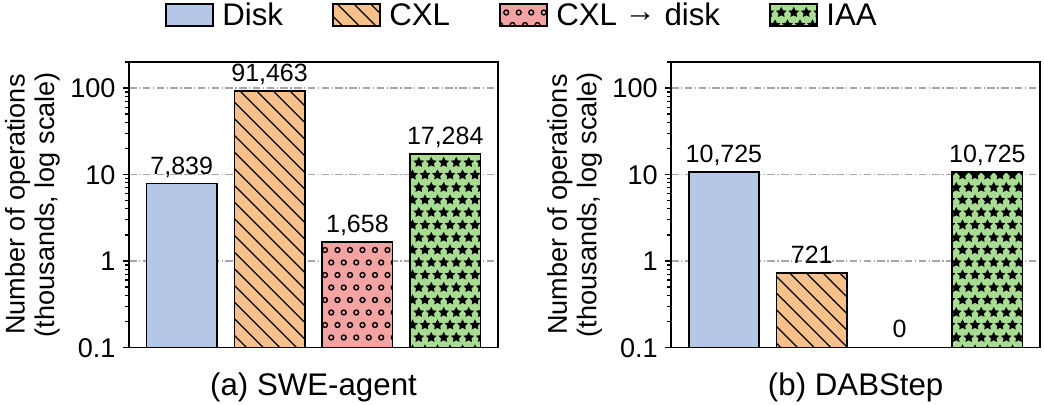}
    \vspace{-5mm}
    \caption{Number of memory-management operations per mechanism with \system{} across workloads.}
    \label{fig:operation-counts}
    \vspace{-3mm}
\end{figure}

\vspace{2pt}
\noindent \textbf{2. Frequency of memory management operations.}
Figure~\ref{fig:operation-counts} shows the number of operations
that \system{} performs with each mechanism during an experiment.
In SWE-agent,
most operations move sandboxes to CXL memory, followed
by IAA compression and SSD offload.
When CXL memory fills up, \system{} also moves sandboxes
that wait for the user from CXL memory to the SSD. In DABStep,
\system{} rarely uses CXL memory, and instead compresses sandboxes
with IAA and moves them to the SSD. 
Overall, \system{} uses all mechanisms, and adapts their mix to each workload.

\begin{figure}[t]
    \vspace{0mm}
    \centering
    \includegraphics[width=\linewidth]{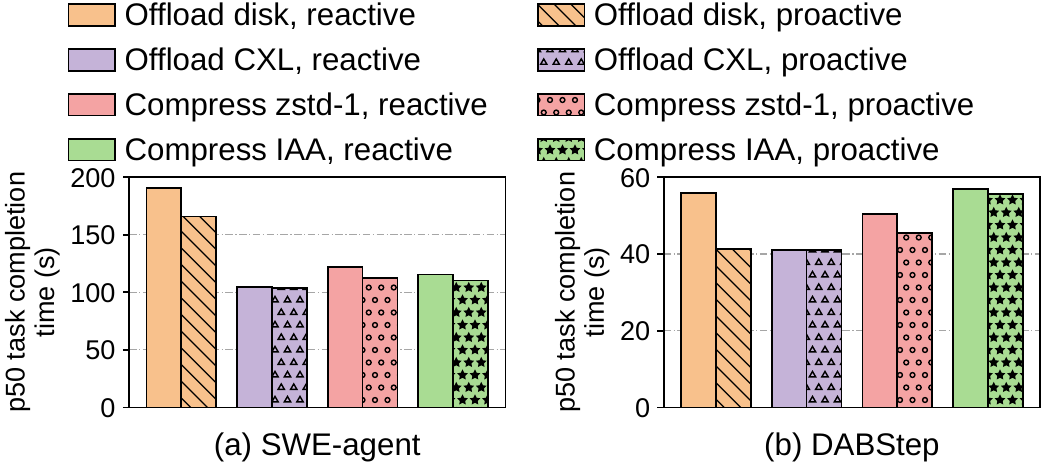}
    \vspace{-5mm}
    \caption{P50 task completion time of the single-mechanism configurations with reactive and proactive restoration.}
    \label{fig:reactive-vs-proactive-p50}
    \vspace{-2mm}
\end{figure}

\vspace{2pt}
\noindent \textbf{3. Reactive vs. proactive sandbox management.}
Figure~\ref{fig:reactive-vs-proactive-p50} compares the
single-mechanism configurations with reactive and proactive
restoration, which starts early to complete when the wait ends.
Proactive restoration reduces the P50 task time by up to 13\% in
SWE-agent and 26\% in DABStep. However, in SWE-agent, proactive
\emph{Offload disk} still takes 70\% longer than \emph{Baseline},
since an SSD restoration often outlasts the LLM
call (\S~\ref{sec:char-tiering}).  \system{} moves a sandbox to the SSD
only when its idle period hides the restoration.

\subsection{Sensitivity Analyses}
\label{sec:sensstudies}

\begin{figure}[t]
    \vspace{0mm}
    \centering
    \includegraphics[width=\linewidth]{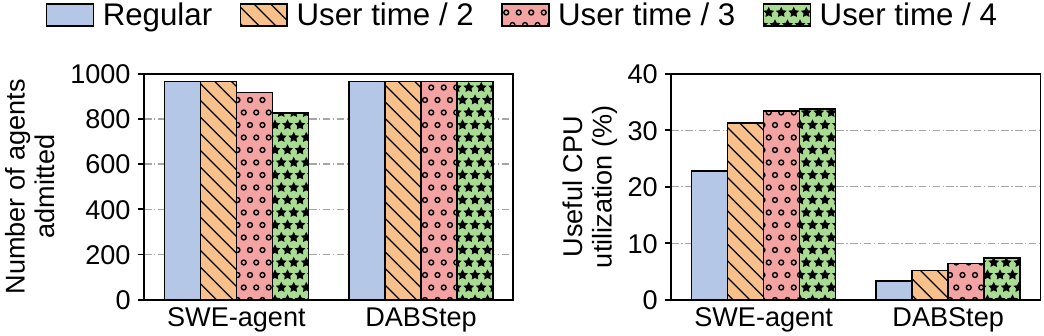}
    \vspace{-5mm}
    \caption{Number of agents admitted and useful CPU utilization with \system{} under different user idle times.}
    \label{fig:sens-user}
    \vspace{-4mm}
\end{figure}

\vspace{2pt}
\noindent \textbf{1. User idle time.}
When users respond faster, 
load on the server increases.
Figure~\ref{fig:sens-user} shows the number of agents that
\system{} admits and the useful CPU utilization when we reduce
the user idle times of the traces by 2, 3, and 4$\times$. In DABStep,
\system{} admits the same number of agents with all idle times. In SWE-agent,
it admits 965 agents with half of the idle time, and 827 agents
with a quarter of it. 
However, each agent executes its tools more often. 
Hence, the
useful CPU utilization increases. With a quarter
of the user idle time, it increases from 22.8\% to 33.8\% in
SWE-agent, and from 3.3\% to 7.5\% in DABStep. The
other metrics, e.g., tool execution times, remain 
constant. Hence, under higher
load, \system{} continues to admit many agents and uses the cores
of the server effectively.

\begin{figure}[t]
    \vspace{0mm}
    \centering
    \includegraphics[width=\linewidth]{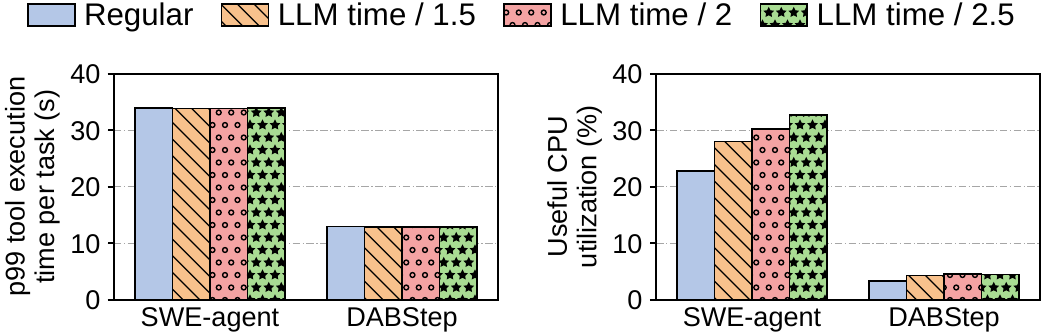}
    \vspace{-5mm}
    \caption{P99 tool execution time per task and useful CPU utilization with \system{} under different LLM inference times.}
    \label{fig:sens-llm}
    \vspace{-2mm}
\end{figure}

\vspace{2pt}
\noindent \textbf{2. LLM inference time.}
Faster LLM inference shortens the idle periods in a task.
Hence, \system{} has less time to reclaim and restore sandbox memory. Figure~\ref{fig:sens-llm} shows P99 total tool
execution time per task and useful CPU utilization when we
divide the LLM inference times by 1.5, 2, and 2.5. 
We can see that P99 tool
execution time stays nearly constant. 
The reason is
that \system{} places a sandbox in a tier only if the sandbox can
return from it before its next tool, i.e., it adapts its placement to the shorter idle periods. 
Further, \system{} admits the same number of agents
with all inference times. However, as agents spend less
time waiting for the LLM, the useful CPU utilization
increases. With
2.5$\times$ faster inference, it increases from 22.8\% to 32.6\%
in SWE-agent, and from 3.3\% to 4.5\% in DABStep. Hence,
\system{} remains effective as LLM inference becomes faster, and
converts the faster inference into higher CPU utilization.

\begin{figure}[t]
    \vspace{0mm}
    \centering
    \includegraphics[width=\linewidth]{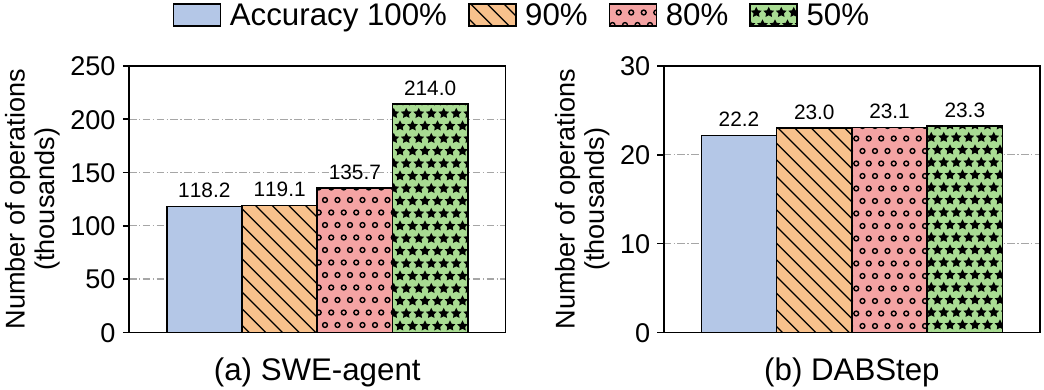}
    \vspace{-5mm}
    \caption{Total number of memory-management operations under different accuracies of the LLM wait time predictor.}
    \label{fig:sens-acc}
    \vspace{-3mm}
\end{figure}

\vspace{2pt}
\noindent \textbf{3. Predictor accuracy.}
\system{} uses its prediction of LLM wait time to select the
placement of each sandbox and the start of its restoration. To
evaluate the effect of mispredictions, we emulate predictors with
100\%, 90\%, 80\%, and 50\% accuracy.
We emulate a mean absolute percentage error of $a\%$ by multiplying each actual LLM wait time by a factor uniformly sampled from $[1-2a/100, 1+2a/100]$.
In all setups, the
number of admitted agents, useful CPU utilization, and task times stay within 1\% of the 100\% accurate predictor. 
However, mispredictions
increase the work of memory management.
Figure~\ref{fig:sens-acc} shows the total number of
memory-management operations, across all the mechanisms of
Figure~\ref{fig:operation-counts}, with each accuracy. As the
accuracy decreases, \system{} moves and (de)compresses sandbox
memory more often. For example, a sandbox that \system{} restores
too early may need to be reclaimed again. With 50\% accuracy,
\system{} performs 81.0\% and 5.0\% more operations than with the
100\% accurate predictor in SWE-agent and DABStep, respectively.
Hence, an accurate predictor keeps the memory-management overhead
of \system{} low, but has a negligible effect on task performance.

\section{Related Work}
\label{sec:related}

\noindent \textbf{Agentic AI characterization.}
Prior work characterized the execution patterns and resource needs of agentic workloads
\cite{kim2026dynamicreasoning,yuan2026workload,zhu2026tracelab,liu2026agenticcoding}. CPU-centric studies identify substantial host-side overhead~\cite{raj2025cpucentric}, while others showed fragmented CPU--GPU execution, bursty resource utilization, and inefficiencies in conventional server architectures~\cite{yang2026architectural}. Our work focuses on the memory behavior of agentic sandboxes, characterizing memory reuse across tool invocations and the overhead of tiering, compression, and restoration. We use these insights to develop a coordinated memory manager that increases agent density while limiting execution overhead.

\vspace{2pt}
\noindent \textbf{Systems for agentic AI.}
Recent works improved efficiency of agentic AI by exploiting workflow dependencies and application knowledge to optimize LLM serving and scheduling~\cite{lin2024parrot,luo2026agentix,kang2026thunderagent,agarwal2026softwaredefined,yu2026pythia}. Others optimized resource allocation and workflow orchestration~\cite{chaudhry2026murakkab}, or used parallelism and speculative execution to reduce the latency of multi-step agentic workflows~\cite{xie2025aimetropolis,ye2026speculativeactions}. These systems primarily optimize agent serving efficiency. Our work focuses on reclaiming CPU memory retained by stateful agentic sandboxes during idle intervals through coordinated memory tiering and compression.

\vspace{2pt}
\noindent \textbf{Memory tiering.}
Systems explored software-defined far memory and transparent memory offloading to reduce DRAM consumption while controlling application performance degradation~\cite{lagarcavilla2019far,weiner2022tmo}. CXL-enabled systems improve page placement across heterogeneous memory tiers through OS-level policies~\cite{maruf2023tpp} and hardware-assisted memory access profiling~\cite{zhou2024neomem}. Others use dynamic page classification and page-size selection~\cite{lee2023memtis}, while reducing migration overhead through asynchronous migration and hardware acceleration~\cite{xiang2024nomad,liu2025dsa2lm}. 
Conventional datacenters use 
memory tiering~\cite{zhong2024cxltiering} and memory harvesting~\cite{fuerst2022memoryharvesting} in virtualized environments. Our work exploits idle intervals in agentic workloads to coordinate memory reclamation (tiering plus compression) and next 
tool execution. 

\vspace{2pt}
\noindent \textbf{Memory compression.}
Researchers characterized compression opportunities in datacenters~\cite{jeong2023compression} and explored hardware acceleration to reduce compression overhead~\cite{abali2020compression,karandikar2023cdpu}. Others improve hardware-compressed memory via efficient data access, address translation, and memory allocation~\cite{chu2026randomaccess,panwar2024dylect,laghari2024memoryallocation}. Prior systems reduced serverless VM memory and restoration overheads via hardware-accelerated snapshot compression~\cite{lazarev2024sabre}, snapshot-based initialization and working-set prefetching~\cite{ustiugov2021reap}, and OS-assisted snapshot restoration~\cite{holmes2026spice}. Our work coordinates memory tiering and compression with agent idle times, selecting mechanisms based on expected idle duration, memory characteristics, and restoration latency to increase memory efficiency while limiting overheads.
\section{Conclusion}
\label{sec:conclusion}

As idle agent sandboxes
keep their memory while they wait for an LLM call or the user, 
datacenter servers run
out of memory before they fully use their cores. We presented \system{}, an agent-aware memory manager that places
each idle sandbox in a local, CXL, or SSD tier, plain or compressed,
that it can return from in time. \system{} reduces the overheads with
efficient snapshotting, hot-page tracking, and hardware acceleration.
Compared to conventional sandbox management, \system{} increases the
number of agents per server by 5.9$\times$ and CPU utilization by 5.0$\times$,
while increasing P50 and P99 latency by up to 5.6\% and 6.3\%, respectively.

\bibliographystyle{ACM-Reference-Format}
\bibliography{refs}

\end{document}